\documentclass[12pt]{article}
\usepackage{latexsym}
\begin{document}
\bibliographystyle{unsrt} 
 
\begin{center} 
{\LARGE \bf Self-Dual Fields and Quaternion Analyticity}\\[10mm] 
 
Sultan Catto$^{\dag}$\\{\it Physics Department\\ The Graduate School and University Center\\ The City University of New York \\365 Fifth Avenue\\
New York, NY 10016-4309\\ and \\ Center for Theoretical Physics \\The Rockefeller University \\
1230 York Avenue\\ New York NY 10021-6399}\\[6mm] 
\end{center} 
\vbox{\vspace{5mm}}

\begin{abstract}
Quaternionic formulation of $D=4$ conformal group and of its associated twistors and their relation to harmonic analyticity is presented. Generalization of $SL(2,{\cal{C}})$ to the $D=4$ conformal group $SO(5,1)$ and its covering group $SL(2,{\cal{Q}})$ that generalizes the euclidean Lorentz group in $R^4$ [namely $SO(3,1)\approx SL(2,{\cal{C}})$ which allow us to obtain the projective twistor space $CP^3$] is shown. Quasi-conformal fields are introduced in $D=4$ and Fueter mappings are shown to map self-dual sector onto itself (and similarly for the anti-self-dual part). Differentiation of Fueter series and various forms of differential operators are shown, establishing the equivalence of Fueter analyticity with twistor and harmonic analyticity. A brief discussion of possible octonion analyticity is provided. 
\end{abstract}
\vbox{\vspace{5mm}}


\vbox{\vspace{10mm}}

$^\dag$ Work supported in part by DOE contracts No. DE-AC-0276 ER 03074 and 03075, and PSC-CUNY Research Awards.
\newpage

\section {Introduction}

We give an expanded review of recent developments of Fueter's theory as discussed in G\"ursey and Tze's book "On the Role of Division, Jordan and Related Algebras in Particle Physics."$^{\cite{gurseyandtze}}$ A brief discussion to possible extension to octonionic analyticity will be presented. An appendix is added to the review on use of coherent states and coset decompositions. 

There are intriguing correspondences between the rotation (or pseudorotation) groups and
conformal groups in $D=1, 2, 4, 6, 10$ Euclidean or Lorentzian space-times due to the properties
of the underlying division algebras. The conformal group is infinite dimensional in $D=2$ and
trivially so in $D=1$. This
has led to the development of conformal field theories and string theories on the $D=2$ world
sheet as well as solvable lattice models in $D=2$ statistical mechanics. These features again
appear in $D=2+1$ topological field theories and quantum gravity. There are also some
correspondences to the self dual and anti self dual sectors of $D=4$ gauge theories. Here we shall
confine ourselves to the cases $D=2, 4$ associated respectively with $\cal{C}$ and $\cal{H}$. The
situation is summarized in the following table:
\vskip 1cm

\begin{center}
\begin{tabular}{|l|c|c|c|}\hline
Dim. & Rot. & Conf. & Div. Alg.  \\ \hline \hline
$D=1$  & $Z_{2}$ &  \shortstack{ $SO(2,1) \sim$\\ $SL(2,{\cal{R}})$} & $\cal{R}$ \\ \hline
$D=1+1$ & $O(1,1)$ & \shortstack{ $SO(2,2) \sim $\\ $SL(2,{\cal{R}})\times SL(2,{\cal{R}})$} &
$\cal{R}\otimes \cal{R}$\\\hline
$D=2$ & \shortstack{$O(2) \sim $ \\ $ U(1)$}  & \shortstack{$SO(3,1) \sim $ \\ $SL(2,{\cal{C}})$} &$\cal{C}$\\\hline
$D=3+1$ & \shortstack{ $SO(3,1) \sim $ \\ $ SL(2,{\cal{C}})$} & \shortstack{$SO(4,2) \sim $ \\ $ SU(2,2)$ } & $\cal{C}$ \\\hline
$D=4$ & \shortstack{$O(4) \sim $ \\ $ SU(2)\times SU(2)$} & \shortstack{$SO(5,1) \sim $ \\ $ SL(2,{\cal{Q}})$} & $\cal{H}$\\\hline 
    \end{tabular}
\end{center}
\vskip 1.5cm

In the Lorentzian two dimensional space the conformal group factorizes, giving rise to left
movers and right movers associated with real wave functions $f_{L}(x+t)$, $f_{R}(x-t)$. On the
other hand $O(4)$ factorizes in euclidean $D=4$ where real three dimensional self dual fields
exist associated with $(1,0)$ and $(0,1)$ representations of $SU(2)\times SU(2)$. In $D=2$
instead of left and right movers we can define analytic and anti analytic functions $f(z)$ and
$g(\bar{z})$, $\overline{f(z)}$ being anti analytic. This is analogous to the $(1,0)$
representation of $O(4)$ in $D=4$ being complex and $(0,1)$ being like the complex conjugate
of $(1,0)$ just like analytic and anti analytic functions. Thus some aspects of analytic functions
in $D=2$ will have their counterparts in the self dual or anti self dual sectors in $D=4$ with
complex numbers being replaced by quaternions. These are the possibilities we shall now explore.

\section{The Conformal group in $D=4$, quaternions and twistors}

The conformal group in Euclidean $D=4$ is the non-compact Lie group $SO(5,1)$ with the
covering group $SL(2,{\cal{Q}})$. These groups generalize the Lie groups $SO(3,1) \sim SL(2,{\cal{C}})$ in Euclidean $D=4$. 

In $D=2$ we can start from the two-dimensional representations of $SL(2,{\cal{C}})$ namely
the complex spinor $\psi_{L}$ with components $\psi^{\alpha}$ ($\alpha=1,2$). It transforms under the group as

\begin{equation}
\psi_{L}^{'} = L~\psi_{L},~~~~~({\rm Det}~L = 1)
\end{equation}
where $L$ is a $2\times2$ complex unimodular matrix, $\psi_{L}$ is a $(\frac{1}{2},0)$
representation. The right handed spinor $\psi_{R}$ associated with the $(0,\frac{1}{2})$
representation transforms as

\begin{equation}
\psi_{R}^{'} = L^{\dag -1}~\psi_{R} \label{1.2}
\end{equation} 

Because of the identity

\begin{equation}
L^{\dag -1} = (i \sigma_{2})^{-1}~ L^{\ast}~i\sigma_{2} = \sigma_{2}~L^{\ast}~\sigma_{2},
\end{equation}
the $CP$ conjugate of $\psi_{L}$, defined as

\begin{equation}
\hat{\psi}_{L} = -i\sigma_{2}~ \psi_{L}^{\ast} = \left(
\begin{array}{c}
-\psi_{2}^{\ast}\\
~\psi_{1}^{\ast}
\end{array}
\right)
\end{equation}
transforms like a right handed spinor $\psi_{R}$ according to the law in Eq.(\ref{1.2}) 

\begin{equation}
\psi_{L}^{'} = L^{\dag-1}~\psi_{L}
\end{equation}

When $SL(2,C)$ is interpreted as the Lie conformal group in $D=2$, $\psi_{L}$ is a twistor
transforming linearly under $SO(3,1)$ which is generated by translations, rotations, dilatation
and inversion. $\psi^{\alpha}$ can be regarded as the homogeneous coordinates of a point in the
projective space $CP^{1}$. The inhomogeneous coordinate is

\begin{equation}
z=\frac{\psi_{1}}{\psi_{2}}   \label{1.6}
\end{equation}
which transforms non linearly under $SL(2,C)$ represented by $L$

\begin{equation}
L = \left(
\begin{array}{cc}
a & b \\
c & d
\end{array}
\right),~~~~~~(ad-bc = 1)
\end{equation}

Then $z$ transfoms undergoes a M\"{o}bius transformation

\begin{equation}
z' = \frac{(az+b)}{(cz+d)},
\end{equation}
with special cases giving:\\
a) dilatation

\begin{equation}
z \longrightarrow k~z ~~~~~~(k= {\rm real,~ positive})
\end{equation}
b) rotation

\begin{equation}
z \longrightarrow e^{i \alpha}~z~~~~~~(\alpha = {\rm real})
\end{equation}
c) translation

\begin{equation}
z \longrightarrow z + b~~~~~~(b = {\rm complex})
\end{equation}
d) inversion

\begin{equation}
z \longrightarrow \frac{1}{z}
\end{equation}
e) special conformal transformation (translation of the inverse)

\begin{equation}
\frac{1}{z} \longrightarrow \frac{1}{z} - s^{\ast}~~~~~~~(s= {\rm complex})
\end{equation}
or

\begin{equation}
z \longrightarrow \frac{z}{1-s^{\ast} z}
\end{equation}

We now consider homogeneous functions of the spinor $\psi_{L}$ ($2-D$ twistor) of the form $F(\psi_{1}, \psi_{2})$ such that

\begin{equation}
F(\lambda \psi_{1}, \lambda \psi_{2}) = \lambda^\delta~F(\psi_{1}, \psi_{2})
\end{equation}
$\lambda$ being complex and $\delta$ denoting the degree of homogeneity.

Choosing $\lambda = \psi_{2}^{-1}$ and using Eq.(\ref{1.6}) we obtain

\begin{equation}
F(z,1) = f(z)= \psi_{2}^{-\delta}~F(\psi_{1},\psi_{2})
\end{equation}
Under $SL(2,C)$ we have

\begin{equation}
\psi_{1}^{'} = a~\psi_{1} + b~\psi_{2},~~~~~~\psi_{2}^{'} = c~\psi_{1} + d~\psi_{2},
\end{equation}
so that we can define

\begin{equation}
f^{'}(z^{'}) = (cz+d)^{\delta}~f(\frac{az+b}{cz+d})
\end{equation}

These correspond to the $(\delta, 0)$ representation of $SL(2,C)$ in terms of homogeneous
analytic functions $f^{(\delta)}(z)$. The $(0,\bar{\delta})$ representations are obtained through
the transformation

\begin{equation}
\psi \longrightarrow \hat{\psi}, ~~~or~~~z \longrightarrow -\frac{1}{z},
\end{equation}
and

\begin{equation}
L= \left(
\begin{array}{cc}
a & b\\
c & d
\end{array}
\right)  \longrightarrow L^{\dag-1}= \left(
\begin{array}{cc}
\bar{d} & -\bar{c}\\
-\bar{b} & \bar{a}
\end{array}
\right),
\end{equation}
so that

\begin{equation}
G^{(\bar{\delta})}(-\bar{\psi}_{2} \bar{\lambda},\bar{\psi}_{1}
\bar{\lambda})=\bar{\lambda}^{\bar{\delta}}~
G^{(\bar{\delta})}(-\bar{\psi}_{2},\bar{\psi}_{1}),
\end{equation}
or, taking $\lambda = \bar{\psi}_{2}^{-1}$

\begin{equation}
G^{(\bar{\delta})}(-\bar{\psi}_{2},\bar{\psi}_{1})=(\bar{\psi}_{2})^{\bar{\delta}}~
G^{(\bar{\delta})}(-1,\bar{z}),
\end{equation}
which gives

\begin{equation}
g(\bar{z}) = G^{(\bar{\delta})}(-1,\bar{z})= (\bar{\psi}_{2})^{-\bar{\delta}}~
G^{(\bar{\delta})}(-\bar{\psi}_{2},\bar{\psi}_{1})
\end{equation}
Hence

\begin{equation}
g^{'}({\bar{z}}^{'})= (\bar{c} \bar{z}+\bar{d})^{\bar{\delta}}~g(\frac{\bar{a}
\bar{z}+\bar{b}}{\bar{c} \bar{z}+\bar{d}})
\end{equation}
For the $(\delta,\bar{\delta})$ representation, we have

\begin{equation}
h^{'(\delta,\bar{\delta})}(z^{'},{\bar{z}}^{'})= (cz+d)^{\delta}(\bar{c}
\bar{z}+\bar{d})^{\bar{\delta}}h^{(\delta,\bar{\delta})}(\frac{az+b}{cz+d},\frac{\bar{a}
\bar{z}+\bar{b}}{\bar{c} \bar{z}+\bar{d}})
\end{equation}
Such primary fields transform like

\begin{equation}
(dz)^{\lambda} (d\bar{z})^{\bar{\lambda}}
\end{equation}

Indeed under

\begin{equation}
z'=\frac{az+b}{cz+d}
\end{equation}
we have

\begin{equation}
dz^{'} = (cz+d)^{-2}~dz   \label{1.28}
\end{equation}
leading to

\begin{equation}
(dz')^{\lambda} (d\bar{z}')^{\bar{\lambda}}= (cz+d)^{-2\lambda}(\bar{c}
\bar{z}+\bar{d})^{-2\bar{\lambda}}~(dz)^{\lambda} (d\bar{z})^{\bar{\lambda}}
\end{equation}
Hence

\begin{equation}
\lambda = - \frac{\delta}{2},~~~~~~\bar{\lambda}= - \frac{\bar{\delta}}{2}
\end{equation}

Under a more general transformation

\begin{equation}
z \longrightarrow w(z),~~~~~~dz\longrightarrow w'(z) dz,    \label{1.31}
\end{equation}
we obtain

\begin{equation}
h^{(\lambda,\bar{\lambda})}(z,\bar{z}) \longrightarrow w'(z)^{\lambda}
\overline{w'(z)}^{\bar{\lambda}} h^{(\lambda,\bar{\lambda})}(w(z),\bar{w}(\bar{z}))
\end{equation}

The transformation (\ref{1.31}) is still a conformal mapping in $D=2$ and corresponds to an
infinite group that generalizes the M\"obius transformation (\ref{1.28}).

We now turn to the generalization of $SL(2,C)$ to the conformal group $SL(2,Q)$ in $D=4$. Its
$2\times 2$ quaternionic representation is

\begin{equation}
\Lambda = \left(
\begin{array}{cc}
a & b\\
c & d
\end{array}
\right), ~~~~~~{\rm Det}~\Lambda =\mid ac^{-1}dc-bc\mid^{2} = 1
\end{equation}
$\Lambda$ acts on a two dimensional quaternionic ket $w$ in quaternionic Hilbert spaces

\begin{equation}
w = \left(
\begin{array}{c}
V\\
\bar{U}
\end{array}
\right),~~~~w' = \Lambda w
\end{equation}
with the $2\times 2$ complex representations

\begin{equation}
V=\left(
\begin{array}{cc}
v_{1} & -v_{2}^{\ast}\\
v_{2} & v_{1}^{\ast}
\end{array}
\right),~~~~~~\bar{U} =  \left(
\begin{array}{cc}
u_{1} & -u_{2}^{\ast}\\
u_{2} & u_{1}^{\ast}
\end{array}
\right)    
\end{equation}
Thus $w$ can be represented by the $4\times 2$ complex matrix with first column $\psi$ and
second column $\hat{\psi}$, where

\begin{equation}
\psi = \left(
\begin{array}{c}
v_{1}\\
v_{2}\\
u_{1}\\
u_{2}
\end{array}
\right),~~~~~~\hat{\psi} = -i \sigma_{2} \psi^{\ast} = \left(
\begin{array}{cc}
 -v_{2}^{\ast}\\
 v_{1}^{\ast}\\
-u_{2}^{\ast}\\
 u_{1}^{\ast}
\end{array}
\right)
\end{equation}
The inverse $SL(2,Q)$ matrix $\Lambda^{-1}$ can also act on a $(1\times 2)$ quaternionic row
from the right, so that

\begin{equation}
s^{\dag} = (\bar{T}, R),~~~~s^{'\dag}= s^{\dag} \Lambda^{-1}     \label{1.37}
\end{equation}

\begin{equation}
\Lambda^{-1}= \left(
\begin{array}{cc}
(a-bd^{-1}c)^{-1} & (c-db^{-1}a)^{-1}\\
(b-ac^{-1}d)^{-1} & (d-ca^{-1}b)^{-1}
\end{array}
\right)    
\end{equation}
where the bar denotes quaternionic conjugation.

The inhomogeneous coordinate in the projective quaternionic space $HP^{1}$ is the ratio of
$V$ and $U$ so that

\begin{equation}
w =
\left(
\begin{array}{c}
x \bar{U} \\
\bar{U}
\end{array}
\right),~~~~~~ x = V \bar{U}^{-1}~~~~~~{\rm or}~~~~~V = x \bar{U}   \label{1.39}
\end{equation}

The $4$-spinor $\psi$ is the twistor representation of the conformal group $SO(5,1)$. We find

\begin{equation}
x= \left(
\begin{array}{cc}
v_{1} & -v_{2}^{\ast}\\
v_{2} & v_{1}^{\ast}
\end{array}
\right)   \left(
\begin{array}{cc}
u_{1}^{\ast} & u_{2}^{\ast}\\
-u_{2} & u_{1}
\end{array}
\right)~\frac{1}{\mid u_{1}\mid^{2}+\mid u_{2}\mid^{2}}
\end{equation}
or

\begin{equation}
x= \left(
\begin{array}{cc}
u_{1}^{\ast} v_{1}+v_{2}^{\ast} u_{2}& v_{1}u_{2}^{\ast}-v_{2}^{\ast}u_{1}\\
v_{2}u_{1}^{\ast}-v_{1}^{\ast}u_{2} & v_{2}u_{2}^{\ast}+v_{1}^{\ast}u_{1}
\end{array}
\right) (\mid u_{1}\mid^{2}+\mid u_{2}\mid^{2})^{-1}
\end{equation}
The twistor $\psi$ takes the form

\begin{equation}
\psi = \left(
\begin{array}{c}
\mid u\mid x_{+}\\
u
\end{array}
\right)= \left(
\begin{array}{c}
xu\\
u
\end{array}
\right),~~~~u = \left(
\begin{array}{c}
u_{1}\\
u_{2}
\end{array}
\right)= U \frac{1+\sigma_{3}}{2}
\end{equation}
where

\begin{equation}
x_{+}=\mid u\mid^{-1} x u = (x_{0}-i \vec{\sigma} \cdot \vec{x}) u \mid u\mid^{-1}
\end{equation}
and

\begin{equation}
\mid u \mid = \sqrt{\mid u_{1}\mid^{2} + \mid u_{2}\mid^{2}}                                    \label{1.43}
\end{equation}

We also have the $O(4)$ invariant quaternion

\begin{eqnarray}
w = \bar{U}^{-1} V& =& \frac{1}{\mid u_{1}\mid^{2}+\mid u_{2}\mid^{2}}        
\left(
\begin{array}{cc}
u_{1}^{\ast} & u_{2}^{\ast}\\
-u_{2} & u_{1}
\end{array}
\right)    
\left(
\begin{array}{cc}
v_{1} & -v_{2}^{\ast}\\
v_{2} & v_{1}^{\ast}
\end{array}
\right)      \nonumber \\
&=&\frac{1}{\mid u_{1}\mid^{2}+\mid u_{2}\mid^{2}} 
\left(
\begin{array}{cc}
u^{\dag} v & u^{\dag}\hat{v}\\
\hat{u}^{\dag}v & v^{\dag}u
\end{array}
\right)
\end{eqnarray}
Under a conformal transformation $\Lambda \in SO(5,1)$ we have

\begin{equation}
V' = a V + b \bar{U} = (a x + b) \bar{U}
\end{equation}

\begin{equation}
\bar{U}^{'} = cV + d\bar{U} = (cx+d)\bar{U}
\end{equation}
where $a$, $b$, $c$, $d$ are the quaternionic elements of $\Lambda$.

Hence, the position quaternion $x$ transforms under the M\"obius transformation

\begin{equation}
x' = (ax+b)(cx+d)^{-1}
\end{equation}
while

\begin{equation}
y' = (cx+d)^{-1}(ax+b)
\end{equation} 

Special cases, together with their infinitesimal forms, are

a) Translations

\begin{equation}
x \longrightarrow x+b~~~~~~~~~(\delta_{\epsilon}x = \epsilon)
\end{equation}

b) Dilatations

\begin{equation}
x \longrightarrow \lambda x,~~~~~({\rm Vec}~\lambda=0, \lambda>0),~~~(\delta_{\kappa}x=\kappa
x)
\end{equation}

c) Left rotations

\begin{equation}
x \longrightarrow m x,~~~~~(\mid m\mid = 1),~~~[\delta_{\mu} x = \mu x, ~~({\rm Sc}~\mu = 0)],
\end{equation}

d) Right rotations

\begin{equation}
x \longrightarrow x \bar{n},~~~~~(\mid n\mid=1),~~~[\delta_{\nu} x = x \nu, ~~({\rm Sc}~\nu = 0)],
\end{equation}

e) Inversion

\begin{equation}
x \longrightarrow x^{-1}, ~~~~~({\rm no~~ infinitesimal~~ form})
\end{equation}

f) Special conformal transformations

\begin{equation}
x^{-1} \longrightarrow x^{-1}- \bar{c}, ~~~{\rm or}~~~x \longrightarrow x(1-\bar{c} x)^{-1},~~
(\delta_{c} x = x \bar{c} x)    \label{53}
\end{equation}

Like in the case of $SL(2,C)$ we consider homogeneous functions of the $SO(5,1)$ twistor.
These can be regarded as functions of $x^{\mu}$ and the spinor $u$, or the spinors $x_{+}$ and
$u$, or again functions of quaternions $V$ (equal to $x\bar{U}$) and $\bar{U}$.

Under the $O(4) \sim SU(2)\times SU(2)$ subgroup of $SL(2,Q)$ we have

\begin{equation}
x' = m x \bar{n},~~~ V \longrightarrow mV,~~~\bar{U} \longrightarrow n\bar{U},  \label{54}
\end{equation}

\begin{equation}
x_{+}^{'} = m x_{+},~~~u' = n u
\end{equation}

We could also introduce another position vector $y$ through the components of the left twistor 
$s^{\dag}$ of Eq. (\ref{1.37})

\begin{equation}
R^{-1} \bar{T} = \bar{y},~~~~~R^{'-1} \bar{T}^{'} = \bar{y}^{'}
\end{equation}

\begin{equation}
(\bar{T}, R) = (R \bar{y}, R)
\end{equation}    

Under the $O(4)$ subgroup we have

\begin{equation}
\Lambda^{-1} = \left(
\begin{array}{cc}
\bar{m} & 0\\
0 & \bar{n}
\end{array}
\right),
\end{equation}
so that

\begin{equation}
\bar{T}' = \bar{T} \bar{m},~~~~~R' = R \bar{n}
\end{equation}
This gives the transformation law

\begin{equation}
\bar{y}' = n \bar{y} \bar{m}, ~~~or~~~y' = m y \bar{n}
\end{equation}
Hence $y$ also transforms like a $4$-vector.

We find

\begin{equation}
\omega_{L} = x \bar{y}, ~~~~\omega_{L}' = m \omega_{L} \bar{m},
\end{equation}

\begin{equation}
\omega_{r} = \bar{y} x, ~~~~\omega_{R}' = n \omega_{R} \bar{n}
\end{equation}

It follows that from a right twistor and a left twistor we can form the two $4$-vectors
$\omega_{L}$ and $\omega_{R}$ which transform respectively like a $(0,0)+(1,0)$ and
$(0,0)+(0,1)$ representations of the $O(4)$ group. It follows that

\begin{equation}
\lambda_{L}= {\rm Vec}~(x \bar{y}) = \frac{1}{2}(x \bar{y}-y \bar{x}) = -i\vec{\sigma}\cdot
\vec{\lambda}_{L}
\end{equation}
is an anti self-dual antisymmetrical tensor. On the other hand

\begin{equation}
{\rm Sc}~(\omega_{L}) = \frac{1}{2}(x \bar{y}+y\bar{x}) = {\rm Sc}~(\omega_{R})
\end{equation}
is an $O(4)$ invariant scalar.

\section{Introduction of quasi-conformal fields in $D=4$. Covariant Fueter mappings}

In analogy to primary conformal fields in $D=2$ that are homogeneous functions of the spinor
representations of $SL(2,C)\sim SO(3,1)$ we shall construct certain functions of the lowest
dimensional representation of $SL(2,Q)\sim SO(5,1)$ in $D=4$.  We have seen that in $D=2$
homogeneous functions of $(\psi_{1}, \psi_{2})$ can be regarded as analytic functions of
$z=\frac{\psi_{1}}{\psi_{2}}$ once a power of $\psi_{2}$ is factored out. In turn, the analytic
function $f(z)$ represents a conformal mapping of the $z$-plane. Let us then start from a
function $F$ of the two quaternionic components $V$ and $\bar{U}$ of the fundamental
representation of $SL(2,Q)$, or, equivalently a function $G$ of the first column $v$ and $u$ of
$V$ and $\bar{U}$ which form the spinorial components of a twistor. Introducing the position
quaternion $x=V\bar{U}^{-1}$ we have

\begin{equation}
F = F(V, \bar{U})= F(x \bar{U}, \bar{U})
\end{equation}
or

\begin{equation}
G=G(v,u) = G(xu,u) = G(x_{+},u),
\end{equation}
where we have used the definition (\ref{1.43}). 

These functions are homogeneous with respect to dilatation if we can write

\begin{equation}
F(\lambda V, \lambda \bar{U}) = \lambda^{k} F(V, \bar{U})= \lambda^{k} F(x
\bar{U},\bar{U})
\end{equation}

\begin{equation}
G(\lambda v, \lambda u) = \lambda^{k}G(v,u)= \lambda^{k} G(x_{+}\mid u\mid,u)
\end{equation}
We can now choose

\begin{equation}
\lambda = \mid U\mid^{-1} = (U \bar{U})^{-\frac{1}{2}}=\mid u\mid^{-1} = (u_{1} u_{1}^{\ast} + u_{2}
u_{2}^{\ast})^{-\frac{1}{2}}
\end{equation}
so that

\begin{equation}
F(V, \bar{U}) = \lambda^{-k} F(V\mid U\mid^{-1}, \bar{U} \mid U\mid^{-1}),
\end{equation}

\begin{equation}
G(v,u) = \lambda^{-\kappa} G(x_{+}, \frac{u}{\mid u\mid})   \label{1.71}
\end{equation}
If we put

\begin{equation}
w=\frac{u}{\mid u\mid},~~~~~\mid w \mid=1,~~~~~ W=\mid U \mid^{-1} U,
\end{equation}
then W is a unit quaternion denoting a point on the sphere $S^{3}$ parametrized by the
normalized spinor $w$, so that

\begin{equation}
F(V,\bar{U}) = \mid U \mid^{k}~F(x\bar{W},\bar{W}),~~~~(W\bar{W}=1),
\end{equation}

\begin{equation}
G(v,u) = \mid u \mid^{k}~G(x_{+},w),~~~~(w^{\dag} w = 1, ~~x_{+}=xw)
\end{equation}

We can go further, to see for what kind of functions the quaternion $W$ represented by a unitary
matrix can be further reduced.

Let

\begin{equation}
W=Z~e^{-i\sigma_{3} \frac{\alpha}{2}}
\end{equation}
Then

\begin{equation}
Z \in (SU(2) / U(1)=S^{2})
\end{equation}
where $U(1)$ is associated with rotations around the third axis. Since the $2\times 2$ matrix
representation for $W$ is

\begin{equation}
W= \left(
\begin{array}{cc}
w_{1} & -w_{2}^{\ast}\\
w_{2} & w_{1}^{\ast}
\end{array}
\right),
\end{equation}
its general parametrization is

\begin{equation}
W = \frac{1}{\sqrt{1+\mid \xi \mid^{2}}}~
\left(
\begin{array}{cc}
1 & -\zeta^{\ast}\\
\zeta & 1
\end{array}
\right)~\left(
\begin{array}{cc}
e^{i \frac{\alpha}{2}} & 0\\
0 & e^{-i\frac{\alpha}{2}}
\end{array}  \right)
\end{equation}
so that the complex number

\begin{equation}
\zeta =\frac{w_{2}}{w_{1}} =\frac{u_{2}}{u_{1}}
\end{equation}
denotes a point on $S^{2}$.

By taking $\lambda$ in Eq.(\ref{1.71}) complex

\begin{equation}
\lambda = u_{1}^{-1}
\end{equation}
we can write

\begin{equation}
G(v,u)=u_{1}^{k} G(x_{+}, \zeta)
\end{equation}
where

\begin{equation}
x_{+}= (x_{0}-i\vec{\sigma}\cdot \vec{x})~
\left(
\begin{array}{c}
1\\
\zeta
\end{array}   \right)~
\frac{1}{\sqrt{1+\mid \zeta \mid^{2}}},
\end{equation}
and $\zeta$ is a point on $S^{2}$. Since in that case the function of $v_{1}, v_{2}, u_{1},
u_{2}$ are reduced to functions of $ v_{1}/u_{1},v_{2}/u_{1}$ and $u_{2}/u_{1}$, we find
the projective space $CP^{3}$ introduced by Penrose and Ward$^{\cite{c1}}$. Such functions can be also 
interpreted as function of $x_{+}$ and a point on $S^{2}$ parametrized by $\zeta$. This is the
method of harmonic analyticity of F. G\"ursey, V. Ogievetsky and their collaborators C. Devchand, M. Evans, W. Jiang, C-H. Tze and others.$^{\cite{c2},\cite{c2a}, \cite{c2b}, \cite{c2c}}$

We shall now present a third category of functions associated with a quaternionic homogeneity
corresponding to the quaternionic projective space $HP^{1}$.

Consider functions $F(V,\bar{U})$ that transform like some representation of $O(4)$ subgroup
of $SO(5,1)$ with $\mu$, $\nu$ being quaternionic, $F$ is assumed to get multipliers 
$\mu (\lambda)$, $\nu (\lambda)$ such that

\begin{equation}
F(V\lambda, \bar{U} \lambda) = \mu (\lambda)~F(V,\bar{U})~\nu (\lambda)
\end{equation}
or

\begin{equation}
F(V,\bar{U})= \mu^{-1} (\lambda)~F(V\lambda,\bar{U} \lambda)~\nu^{-1}(\lambda)
\end{equation}

Taking

\begin{equation}
\lambda = \bar{U}^{-1}
\end{equation}
we obtain

\begin{equation}
F(V,\bar{U})= \mu^{-1} (\bar{U}^{-1})~F(x,1)~\nu^{-1}(\bar{U}^{-1})
\end{equation}
where we have used Eq.(\ref{1.39}).

More generally $V$ and $\bar{U}$ can also undergo an $O(4)$ transformation

\begin{equation}
V \longrightarrow m V,~~~~ \bar{U} \longrightarrow n\bar{U},~~~~~ (\mid m \mid=\mid n \mid=1)
\end{equation}
Hence, we can have more general multipliers $M$ and $N$. Indeed,

\begin{equation}
F(m V \lambda, n \bar{U} \lambda) = F(m V \lambda, m p \bar{U} \lambda)
\end{equation}
where have have put

\begin{equation}
n=mp
\end{equation}
Taking

\begin{equation}
\lambda = \bar{U}^{-1} p^{-1} \bar{m},~~~~~(\bar{m}=m^{-1})
\end{equation}
we find

\begin{eqnarray}
F(V,\bar{U}) &=& M^{-1}~F(mV\lambda,mp\bar{U} \lambda)~N^{-1} \nonumber \\
&=& M^{-1}~F(mxp^{-1} \bar{m})~N^{-1}
\end{eqnarray}
where $M$ and $N$ are in general function of $m$, $p$ and $\bar{U}$.  If $F$ is a scalar, 
then $M$ and $N$ can not depend on $m$ and $n$, or alternatively $m$ and $p$.  They will depend on the scalar $\mid U \mid$.  Hence in that case if the homogeneity degree is $k$ we have

\begin{equation}
F(V,\bar{U})= \mid U \mid^{k}~F(mxp^{-1}\bar{m})
\end{equation}

If $F$ is a quaternion transforming like $(0,0) + (1,0)$ under $O(4)$, then we must have

\begin{equation}
F^{(k)}(V,\bar{U})=\mid U \mid^{k}~m^{-1}~F^{(k)}(mxp^{-1}\bar{m})m
\end{equation}

Such a function is a power series in

\begin{equation}
Z=xp^{-1}
\end{equation}
so that

\begin{equation}
F_{L}(z)=\sum_{n} c_{n} Z^{n}~~~~~~~(Vec~c_{n}=0)
\end{equation}

These are just left-right holomorphic Fueter functions of $z$. If $F_{R}$ transform like $[(0,0)
+ (0,1)]$ representation of $O(4)$ then we have a function of

\begin{equation} 
S=p^{-1} x
\end{equation}
where $p$ transforms like a $4$-vector under $O(4)$ as

\begin{equation}
p \longrightarrow mp\bar{n}
\end{equation}
so that

\begin{equation}
S \longrightarrow n S \bar{n}
\end{equation}
Then

\begin{equation}
F_{R}^{(k)}(V,\bar{U})=\mid U \mid^{k}~n^{-1}~F^{(k)}(np^{-1}x\bar{n})~n
\end{equation}
If $F$ transforms like a $4$-vector, then it is of the form

\begin{equation}
F=p^{-1}~\sum_{n} c_{n} (xp^{-1})^{n} = p^{-1}~F_{L}(Z)=F_{R}(S)p^{-1}
\end{equation}
It transforms as

\begin{equation}
F \longrightarrow n~F~\bar{m}
\end{equation}
which is the transformation law for $x^{-1}$.  Thus we have the mapping defined by

\begin{equation}
x^{'-1} = n~(p^{-1}~\sum_{n} c_{n}(xp^{-1})^{n}+\ell^{-1})^{-1}~\bar{m}    \label{101}
\end{equation}
where we have also used an arbitrary $O(4)$ transformation.  Note that unlike the $D=2$ case
this mapping is not conformal.  We shall call it quasi conformal.

Consider the special case

\begin{equation}
c_{n}=-c
\end{equation}
with $n$ going from zero to infinity.  Then the mapping takes the form

\begin{equation}
x^{'} = m~(\frac{c}{x-p}+\ell^{-1})^{-1}~\bar{n}   \label{103}
\end{equation}
which is equivalent to a quaternionic M\"obius transformation that represents a conformal
transformation.  Hence the infinite parameter mapping of Eq.(\ref{101}) admits the conformal
group as a subgroup.

This is the generalization to $D=4$ of the infinite parameter 
holomorphic mapping in $D=2$,

\begin{equation}
z'=f(z)=\sum_{n} b_{n}z^{n}
\end{equation}
where $b_{n},z^{n}$ are complex.  The mapping admits as a finite parameter subgroup

\begin{equation} 
z'=\frac{az+b}{cz+d},~~~~~ (ad-bc=1)
\end{equation}
which is the M\"obius transformation that is a nonlinear realization of $SO(3,1) \sim SL(2,C)$
on the coset 

\begin{equation}
SL(2,C)/\Delta \times E_{2}
\end{equation}
$\Delta$ being the dilatation and $E_{2}$ the Euclidean group in $D=2$.

In the case of $D=2$ the holomorphic group and its M\"obius subgroup are both conformal,
since

\begin{equation}
ds^{2}=dz'~d\bar{z}^{'} = \mid f'(z)\mid^{2}~dz~d\bar{z}
\end{equation}
This is not true in $D=4$.  The infinite group Eq.(\ref{103}) does not lead to a conformally flat
metric, but its M\"obius subgroup does. Indeed we have

\begin{equation}
dx'=-m~(\frac{c}{x-p}+\ell^{-1})^{-1}~dx~(\frac{c}{x-p}+ \ell^{-1})^{-1}~\bar{n}
\end{equation}
so that

\begin{equation}
dx'~d\bar{x}^{'}= \mid (\frac{c}{x-p}+\ell^{-1}) \mid^{-2}~dx~d\bar{x}
\end{equation}

The quaternionic holomorphic mapping of Eq.(\ref{103}) can be called quasi conformal.  If we
make the $O(4)$ transformation

\begin{equation}
X=mx'\bar{n},~~~ x'=mx\bar{n},~~~ P=mp\bar{n},~~~ L=m\ell \bar{n}   \label{110}
\end{equation}
it takes the form

\begin{equation}
X^{'-1} = \sum_{n} c_{n} (P^{-1} x)^{n}~P^{-1} +L^{-1}
\end{equation}
or, putting

\begin{equation}
Z'=P^{-1}X',~~~~~Z=P^{-1} X = Z_{0}-i \vec{\sigma}\cdot \vec{Z}
\end{equation}

\begin{equation}
Z^{'-1}=\sum_{n} c_{n} Z^{n} + L^{-1} P
\end{equation}
In this last form the mapping is recognized as being a left-right 
holomorphic Fueter transformation.

If we write

\begin{equation}
ds^{2} = dZ'~d\bar{Z}^{'} = g_{\mu \nu}~dZ^{\mu}~dZ^{\nu}
\end{equation}
then it is easily shown that in $4$-dimensional polar coordinates we have$^{\cite{c3}}$

\begin{equation}
g_{0n}=0,~~~~~~ (n=1,2,3)
\end{equation}

In fact, putting

\begin{equation}
z=Z_{0} + i\mid \vec{Z} \mid
\end{equation}
it takes the form$^{\cite{c4}}$

\begin{equation}
ds^{2} = 
\Phi^{2}(z,\bar{z})~dz~d\bar{z}+\rho^{2}(z,\bar{z})~d\Omega^{2}
(\theta,\phi),
\end{equation}
where $d\Omega^{2}$ is the line element on $S^{2}$.  Then, this 
Kruskal form 
that generalizes the conformal metric merits the name: quasi conformal.

Another similarity with $D=2$ is provided by differential equations 
satisfied by the mappings.

In $D=2$ we have

\begin{equation}
(\partial_{x}+i\partial_{y})~f(z) =\frac{\partial}{\partial\bar{z}}~f(z)=0   \label{118}
\end{equation}
which imply the harmonicity of $f(z)$, i.e.:

\begin{equation}
\Delta f(z)=\frac{\partial}{\partial z} ~\frac{\partial}{\partial\bar{z}}f(z)=0
\end{equation}
In $D=4$ the mapping

\begin{equation}
x'=F(x)
\end{equation}
defined by Eq.(\ref{101}) satisfies

\begin{equation}
\Box \Box F(x)=0
\end{equation}
showing that $F(x)$ is biharmonic.  We have seen that in a special $O(4)$ 
frame this mapping takes the Fueter form

\begin{equation}
Z'=\sum_{n} c_{n} Z^{n} =\Phi(Z)
\end{equation}
which satisfies

\begin{equation}
\Box D \Phi(z)= D\bar{D} D \Phi(z)=0
\end{equation}
where

\begin{equation}
D= \frac{\partial}{\partial z^{0}}-i \vec{\sigma} \cdot
\frac{\partial}{\partial \vec{z}}~~~~~~\bar{D}= \frac{\partial}{\partial z^{0}}+
i \vec{\sigma} \cdot \frac{\partial}{\partial \vec{z}}     \label{124}
\end{equation}
In other words the function

\begin{equation}
G(z^{0},\vec{z})= \Box \Phi(Z)
\end{equation}
satisfies

\begin{equation}
DG=0
\end{equation}
which is equivalent to the massless $2$-component Dirac equation

\begin{equation}
(\partial_{0} - i \vec{\sigma}\cdot \vec{\nabla}) 
\left(
\begin{array}{c}
g_{1}\\
g_{2}  
\end{array}  
\right)=0                 \label{127}
\end{equation}
where we have used the following $2\times 2$ representation for G:

\begin{equation}
G= \left(
\begin{array}{cc}
g_{1} & -g_{2}^{\ast}\\
g_{2} & g_{1}^{\ast}
\end{array}
\right)
\end{equation}
The Eq.(\ref{127}) generalizes the Cauchy-Riemann Eq.(\ref{118}).

Here we must note that instead of $Z$ standing for $P^{-1}X$ which 
transforms like $(0,0) + (0,1)$ representation of $O(4)$ we could have taken $U$ representing
$XP^{-1}$.  Also writing $U'$ for $X'P^{-1}$ that transforms like 
$(0,0) + (1,0)$ we have an alternative form of the pseudo conformal mapping that reads

\begin{equation}
U'=\sum_{n} c_{n} U^{n}
\end{equation}
Under the $O(4)$ transformation Eq.(\ref{110}) we have

\begin{equation}
Z \longrightarrow nZ\bar{n},~~~~~~Z' \longrightarrow nZ'\bar{n},
\end{equation}

\begin{equation}
U \longrightarrow m U \bar{m},~~~~~~U' \longrightarrow m U' \bar{m}
\end{equation}
This shows that the self-dual sector is mapped into itself, the same being valid for the
anti self-dual sector.

\section{ Geometric Interpretation - Functions on elements of the coset $SO(5,1)/SO(4)\times O(1,1)$}

The covariant Fueter functions can be shown to be elements of a 
function of a coset element 

\begin{equation}
\phi = SL(2,Q)/Sp(1)\times Sp(1)\times O(1,1)
\end{equation}
 represented by a $2\times 2$ quaternionic matrix.  The dilation $O(1,1)$ and the $O(4)\sim Sp(1)\times Sp(1)$ groups act linearly on $\phi$.  We parametrize the $SL(2,Q)$ matrix $\Lambda$ in the following way:

\begin{eqnarray}
\Lambda &=& \left(
\begin{array}{cc}
a & b \\
c & d
\end{array}
\right) \nonumber  \\
&=&  \left(
\begin{array}{cc}
1 & x \\
0 & 1
\end{array}
\right)~ \left(
\begin{array}{cc}
1 & 0 \\
p^{-1} & 1
\end{array}
\right)~ \left(
\begin{array}{cc}
\kappa^{\frac{1}{2}} m & 0 \\
0  & \kappa^{-\frac{1}{2}} n
\end{array}
\right)                            \label{132}
\end{eqnarray}   
or

\begin{equation}
\Lambda=F~H
\end{equation}
where

\begin{equation}
H= \left(
\begin{array}{cc}
\kappa^{\frac{1}{2}} m & 0 \\
0  & \kappa^{-\frac{1}{2}} n
\end{array}
\right),~~~~~~(\kappa~~{\rm real}, \mid m \mid=\mid n \mid=1)
\end{equation}
represents the subgroup $O(4)\times O(1,1)$ and

\begin{equation}
F=\left(
\begin{array}{cc}
1+xp^{-1} & x \\
p^{-1}  & 1
\end{array}
\right),~~~~~~({\rm Det}~F=1)
\end{equation}
is the coset element.  In order to eliminate the subgroup $H$ we introduce the element

\begin{equation}
\eta=\left(
\begin{array}{cc}
1 & 0 \\
0  & 1
\end{array}
\right),~~~~\eta^{2}=I,~~~~[H,\eta]=0,
\end{equation}
which commutes with $H$ and form the matrix

\begin{equation}
\phi = \Lambda \eta \Lambda^{-1} \eta^{-1}=F \eta F^{-1} \eta,~~~~~~
({\rm Det}~\phi=1)
\end{equation}

We have

\begin{equation}
F^{-1}= \left(
\begin{array}{cc}
1 & -x \\
-p^{-1} & 1+p^{-1} x
\end{array}
\right)
\end{equation}
Hence $\phi$ has the form

\begin{eqnarray}
\phi &=& \left(
\begin{array}{cc}
1+v & x \\
p^{-1} & 1
\end{array} \right)
~\left(
\begin{array}{cc}
1 & x \\
p^{-1} & 1+u
\end{array} \right)  \nonumber \\
& =& \left(
\begin{array}{cc}
1+2v & 2(1+v)x \\
2p^{-1} & 1+2u
\end{array} \right)
\end{eqnarray}
where

\begin{equation}
u=p^{-1}x,~~~~~~~v=xp^{-1}
\end{equation}

\begin{equation}
xu=vx=xp^{-1}x
\end{equation}

We can also write

\begin{equation}
\phi = 1 + 2W
\end{equation}
where

\begin{equation}
W(p,x)= \left(
\begin{array}{cc}
v & (1+v)x \\
p^{-1} & u
\end{array} \right)=\left(
\begin{array}{cc}
v & v(1+v)p \\
p^{-1} & u
\end{array} 
\right)
\end{equation}
Under the subgroup $R \in H$ acting on $M$ from the left, we have

\begin{equation}
M \longrightarrow RM,~~~~\phi \longrightarrow R \phi R^{-1},~~~~
W \longrightarrow R W R^{-1}
\end{equation}
or

\begin{equation}
u \longrightarrow nu\bar{n},~~~~~~~~~v \longrightarrow mv\bar{m},
\end{equation}

\begin{equation}
x \longrightarrow \kappa mx\bar{n},~~~~~~
p \longrightarrow \kappa mp\bar{n}
\end{equation}

Consider now a power series in $\phi$, or a function of $W$

\begin{equation}
G(W)=\sum_{n} c_{n} W^{n}
\end{equation}
under $R$ we have

\begin{equation}
G(W) \longrightarrow RG(W)R^{-1}
\end{equation}
The function $G(W)$ is biharmonic when $G(W)$ is regarded as a function of $x$

\begin{equation}
\Box_{x} \Box_{x}G(W(x))=0
\end{equation}
since we have

\begin{equation}
G^{(n)}=W^{n}(x)=\left(
\begin{array}{cc}
G_{11}(v) & h^{(n)}(v) p \\
k^{(n)}(u)p^{-1} & g_{22}(u)
\end{array} \right)
\end{equation}

Introduce the $2\times 2$ Dirac operators

\begin{equation}
D_{x} = \partial_{0} - i \vec{\sigma}\cdot \vec{\nabla},~~~~~
D_{u}=p D_{x},~~~~~D_{v}=D_{x} p
\end{equation}
We have

\begin{equation}
D_{u} \Box G_{22}(u)=0,~~~~or~~~D_{x} \Box_{x} G_{22}(p^{-1}x)=0,
\end{equation}

\begin{equation}
\Box G_{11}(v){\overleftarrow{D}}_{v}=0,~~~~or~~~\Box_{x} G_{11}(xp^{-1}){\overleftarrow{D}}_{x}=0,
\end{equation}

\begin{equation}
D_{v} \Box G_{11}(v)=0,~~~~\Box G_{22}(u) {\overleftarrow{D}}_{v}=0
\end{equation}
Hence, if we define

\begin{equation}
{\cal{D}}=\left(
\begin{array}{cc}
D_{u} & 0 \\
0 & D_{v}
\end{array} \right)
\end{equation}
we have

\begin{equation}
{\cal{D}} \Box G(W)= {\cal{D}} {\bar{{\cal{D}}}} {\cal{D}} G(W)=0,
\end{equation}
which is the condition of Fueter analyticity for the function

\begin{equation}
g(x,p)= \Box G(W)
\end{equation}

The elements of $G$ have definite tensorial properties under $O(4)$:

\begin{equation}
G_{11} \sim (0,0) + (1,0),~~~~~ G_{22} \sim (0,0) + (0,1)
\end{equation}

\begin{equation}
G_{12} \sim (\frac{1}{2},\frac{1}{2}),~~~~~G_{21} \sim 
(\frac{1}{2},\frac{1}{2})
\end{equation}

We now turn to the transformation properties under the remaining
transformations of $SO(5,1)/SO(4)\times O(1,1)$ under which both $x$ and $p$ transform non
linearly.

We have, from Eq.(\ref{132})

\begin{equation}
\Lambda = \left(
\begin{array}{cc}
a & b \\
c & d
\end{array}  \right) =\left(
\begin{array}{cc}
(1+x p^{-1})\kappa^{\frac{1}{2}} m & x\kappa^{-\frac{1}{2}} n \\
p^{-1}\kappa^{\frac{1}{2}} m & \kappa^{-\frac{1}{2}}n
\end{array} \right)
\end{equation}

Under left action of $SL(2,Q)$ we have

\begin{equation}
\Lambda^{'} = K \Lambda,
\end{equation}
where

\begin{equation}
K=\left(
\begin{array}{cc}
\alpha & \beta \\
\gamma & \delta
\end{array} \right),~~~~~~(Det~K=1)
\end{equation}

Solving for $x$ and $p$ we find

\begin{equation}
x=bd^{-1},~~~~~~ p+x = ac^{-1}
\end{equation}
Hence

\begin{equation}
x'=b'd^{'-1},~~~~~~ p'=a'c^{'-1}-b'd^{'-1}
\end{equation}
giving

\begin{equation}
x'=(\alpha x+\beta)(\gamma x+\delta)^{-1},
\end{equation}

\begin{eqnarray}
p'&=& [\alpha(p+x)+\beta][\gamma(p+x)+\delta]^{-1} \nonumber  \\
&~~~~~-& (\alpha x + \beta)(\gamma x + \delta)^{-1}
\end{eqnarray}
It follows that $p$ transforms like the difference of two position vectors, or like $\Delta x$. This
implies that under the translation subgroup

\begin{equation}
x'=x+\beta
\end{equation}

\begin{equation}
p'=p
\end{equation}

Hence $p$ transforms like a momentum under the Poincar\'e subgroup of the conformal group.

We note that

\begin{equation}
\left( \begin{array}{c}
b\\ d
\end{array} \right)
\end{equation}
and

\begin{equation}
\left( \begin{array}{c}
a\\ c
\end{array} \right)
\end{equation}
behave like quaternionic twistors with respect to left multiplication of $\Lambda$ by $K \in SL(2,Q)$

As before we can introduce functions of $x$ and $p$ that acquire 
quaternionic multipliers under an $SL(2,Q)$ transformations.  They are generalized
quasi-conformal fields as discussed in the previous section.

\section{The Quadratic Schwarz Differentials in $D=4$. Differentiation of Fueter Series}

In the case of the complex M\"obius transformation

\begin{equation}
w=\frac{az+b}{cz+d}
\end{equation}
the Schwarz derivative

\begin{equation}
\{w,z\}= \frac{d^{3}w}{dz^{3}} (\frac{dw}{dz})^{-1} - \frac{3}{2}[\frac{d^{2}w}{dz^{2}}
(\frac{dw}{dz})^{-1}]^{2}
\end{equation}
vanishes, so that

\begin{equation}
\{\frac{az+b}{cz+d},z\}=0        \label{171}
\end{equation}

We can also define the quadratic Schwarz differential

\begin{equation}
w(w,z)=\{w,z\}dz^{2}=d^{3}w(dw)^{-1}-\frac{3}{2}[d^{2}w (dw)^{-1}]^{2}
\end{equation}
which also vanishes for the M\"obius transformation.

The Schwarz differential can also be obtained as the limit of the 
cross-ratio of $4$ points when they all tend to a common point.

Consider the successive mappings

\begin{equation}
Z \longrightarrow u(z) \longrightarrow w(u(z))
\end{equation}
Then

\begin{equation} 
\omega (w,z)=\omega (w,u)+\omega (u,z)           \label{174}
\end{equation}
which is equivalent to, but simpler then the well known rule

\begin{equation}
\{w,z\} = \{w,u\}~(\frac{du}{dz})^{2} + \{u,z\}
\end{equation}
In the special case 

\begin{equation}
u=\frac{az+b}{cz+d}
\end{equation}
we have

\begin{equation}
\omega (w,z)=\omega (w,\frac{az+b}{cz+d})
\end{equation}
or

\begin{equation}
\{w,z\} =\{w,\frac{az+b}{cz+d}\}~\frac{1}{(cz+d)^{4}}
\end{equation}

Thus, under $SL(2,C), \{w,z\}$ transforms like $(dz)^{2}$ and $\{w,z\}^{\frac{k}{2}}$ 
like $dz^{k}$ so that it has conformal weight $k$.  However, under $z \rightarrow u(z)$ the
conformal field transforms inhomogeneously, acquiring an additional piece that corresponds to
the central extension of the infinite group of homogeneous analytic transformations.

We now turn to the $D=4$ case and the quasi-holomorphic mappings $x \rightarrow x'$ of
quaternions.  Is is much easier to generalize differentials than derivatives in the non
commutative case.  Hence we define two kinds of quadratic Schwarz differentials, both being
cross ratios of four nearby points in euclidean space-time.

\begin{equation}
\Omega_{L}(w,x) = d^{3}w~dw^{-1} -\frac{3}{2} (d^{2}w~dw^{-1})^{2},
\end{equation}

\begin{equation}
\Omega_{R}(w,x) =dw^{-1}~d^{3}w -\frac{3}{2} (dw^{-1} ~d^{2}w)^{2}
\end{equation}

If $x$ and $w$ are $4$-vectors, then

\begin{equation}
\Omega = Sc~\Omega_{L} = Sc~\Omega_{R}
\end{equation}
is $O(4)$ invariant, while $Vec~\Omega_{L}$ and $Vec~\Omega_{R}$ transform respectively
like a self dual and an anti self dual tensor. Now we have

\begin{equation}
\Omega_{L}~ [(ax+b)(cx+d)^{-1},x]=0,
\end{equation}

\begin{equation}
\Omega_{R}~ [(ax+b)(cx+d)^{-1},x]=0,
\end{equation}
for a quaternionic M\"obius transformation, generalizing Eq.(\ref{171}).

In the case of the more general pseudo-conformal mapping, we seek a generalization of
Eq.(\ref{174}). To this end we introduce the infinite matrix $S$ with elements equal to unity in
the upper line parallel to the diagonal, with all other elements being zero.

\begin{equation}
S= \left(
\begin{array}{cccccc}
0 & 1 & 0 & 0 & 0 & \rightarrow \\
0 & 0 & 1 & 0 & 0 &  \\
0 & 0 & 0 & 1 & 0 &  \\
0 & 0 & 0 & 0 & 1 &  \\
0 & 0 & 0 & 0 & 0 &  \\
\downarrow &  &  &  &  & \ddots
\end{array}
\right)
\end{equation}

We now consider the quaternionic matrix

\begin{eqnarray}
\frac{1}{1-SZ} &=& 1+SZ+SZ^{2} +\cdots \nonumber  \\
&=& \left(
\begin{array}{cccccc}
1 & Z & Z^{2} & Z^{3} & Z^{4} & \cdots  \\
  & 1 & Z & Z^{2} & Z^{3} & \cdots  \\
  &   & 1 & Z & Z^{2} & \cdots \\
  &   &   & 1 & Z  & \cdots \\
  &   &   &   & 1 & \cdots \\
  &   &   &   & \vdots  & \ddots
\end{array}
\right)
\end{eqnarray}

Define the kets

\begin{equation}
\mid \alpha> = \left(
\begin{array}{c}
\alpha_{0} \\
\alpha_{1} \\
\alpha_{2} \\
\alpha_{3} \\
\vdots
\end{array}
\right),~~~~~~~\mid 0> = \left(
\begin{array}{c}
1 \\
0 \\
0 \\
0 \\
\vdots
\end{array}
\right)
\end{equation}
The basis for this kets is provided by the harmonic oscillator operators $a$, ~$a^{\dag}$ which obey 

\begin{equation}
[a, a^{\dag}] = 1
\end{equation}
Then we have

\begin{equation}
\left(
\begin{array}{c}
0 \\
1 \\
0 \\
0 \\
\vdots
\end{array}
\right)= a^{\dag}~\mid 0>=\mid 1>,~~~~\left(
\begin{array}{c}
0 \\
0 \\
1 \\
0 \\
\vdots
\end{array}
\right)= \frac{a^{\dag^{2}}}{\sqrt{2!}}~\mid 0> = \mid 2>
\end{equation}
etc. With these notations it is easy to check that

\begin{equation}
S= \sum_{n=0}^{\infty} \mid n><n+1 \mid = \sum_{n} \frac{a^{\dag^{n}}}{\sqrt{n!}}
\mid 0><0\mid \frac{a^{n+1}}{\sqrt{(n+1)!}}
\end{equation}

\begin{equation}
U= <0\mid (1-SZ)^{-1}\mid \alpha>=\sum_{n} \alpha_{n} Z^{n}= F(Z)
\end{equation}
displaying the Fueter mapping as a matrix element of the operator $(1-SZ)^{-1}$. Its other
matrix elements are also Fueter functions. We can now calculate $dU$:

\begin{equation}
dU= <0\mid (1-SZ)^{-1} S~dZ~(1-SZ)^{-1} \mid \alpha>
\end{equation}

If we have

\begin{equation}
W=W(U), ~~~~~~U=U(Z),
\end{equation}
we can write

\begin{equation}
dW = <0 \mid  \frac{1}{1-SU}~ S~dU~\frac{1}{1-SU} \mid \beta>,
\end{equation}
\begin{eqnarray}
d^{2}W = <0\mid \{ \frac{2}{1-SU}~ S~dU~\frac{1}{1-SU}~ S~dU~\frac{1}{1-SU} \nonumber \\
~~~~~~~~+ \frac{1}{1-SU}~S~d^{2}U~\frac{1}{1-SU} \} \mid \beta>, ~~~~~{\rm etc.}
\end{eqnarray}
while

\begin{equation}
d^{2}U = 2~<0\mid  \frac{1}{1-SZ}~ S~dZ~\frac{1}{1-SZ}~S~dZ~\frac{1}{1-SZ}\mid \alpha>
\end{equation}

These formulae allow us to evaluate quaternionic quadratic Schwarz differentials for Fueter mappings.

Note that the operator

\begin{equation}
H(Z) = (1-SZ)^{-1}
\end{equation}
satisfies

\begin{equation}
\Omega_{L}~(H(Z),Z)=0,~~~~~~\Omega_{R}~(H(Z),Z)=0
\end{equation}
while this is not true for the matrix elements of $H(Z)$. Thus

\begin{equation}
\Omega_{L}~(<0\mid H(Z)\mid \alpha>,Z) \neq 0
\end{equation}

The possibility now arises to define a left or right pseudo-conformal field as a quaternionic
Fueter function that transforms like a power of $\Omega_{L}$ or $\Omega_{R}$. Due to the
inhomogeneous term in the transformation that vanishes for M\"obius transformation such fields
will represent a centrally extended Fueter algebra, the extension being quaternionic in general.
The determination of the exact nature of the extension is still an open problem. The general
problem first investigated by Gelfand and Fuchs$^{\cite{c5}}$ has been recently discussed by Cardy$^{\cite{c6}}$, Fradkin and his collaborators$^{\cite{c7}}$, Nair and Schiff$^{\cite{c8}}$ and others from various angles (see forexample references $^{\cite{c1a}}$, $^{\cite{c1b}}$, $^{\cite{c1c}}$, $^{\cite{c1d}}$, $^{\cite{c1e}}$,$^{\cite{c1f}}$). Many of the formulas of ADHM construction of solutions to self-dual Yang-Mills are most conveniently written in terms of quaternions$^{\cite{Martinec}, \cite{siegel}}$.

\section{Basis of Fueter analytic functions. Relation to harmonic analyticity.}

We have seen that the left-right holomorphic quaternionic Fueter series of the form

\begin{equation}
F(x)= \sum_{n} a_{n} x^{n},~~~~~~({\rm Vec}~a_{n}=0)
\end{equation}
cannot be used in physics if $x$ is a $4$-vector transforming under $O(4)$ as in Eq.(\ref{54})
since powers of $x$ transform like tensor elements, so that $F$ will not have a definite tensorial
property. However functions of $Z$ or $U$, where

\begin{equation}
Z=xp^{-1},~~~~~~U=p^{-1}x
\end{equation}
with $a_{n},~b_{n}$ being $O(4)$ scalars lead to

\begin{equation}
F_{L}(Z)= \sum_{n} a_{n} Z^{n},~~~~~~F_{R}(U)=\sum_{n} b_{n} U^{n}
\end{equation}
such that ${\rm Sc}~F_{L}, ~{\rm Sc}~F_{R}$ are $O(4)$ scalars, $ {\rm Vec}~F_{L}$ is self-dual and
${\rm Vec}~F_{R}$ is anti self-dual.

Then the functions $G_{L},~G_{R}$ defined by

\begin{equation}
G_{L}=\Box F_{L},~~~~~~G_{R}=\Box F_{R}
\end{equation}
are called left-right analytic by Fueter. Under $O(4)$ we have, using the definition
Eq.(\ref{124})

\begin{equation}
G_{L} \longrightarrow m G_{L} \bar{m},~~~~~~G_{R} \longrightarrow nG_{R}\bar{n},
\end{equation}

\begin{equation}
D \longrightarrow m D \bar{n},~~~~~~\bar{D} \longrightarrow n\bar{D} \bar{m}
\end{equation}

Thus we have

\begin{equation}
\bar{D}G_{L} \longrightarrow n \bar{D} G_{L} \bar{m},~~~~~~
G_{L}\overleftarrow{D} \longrightarrow m G_{L} \overleftarrow{D}\bar{n}
\end{equation}

\begin{equation}
DG_{R} \longrightarrow m D G_{R} \bar{n},~~~~~~
G_{R}\overleftarrow{\bar{D}} \longrightarrow n G_{R} \overleftarrow{\bar{D}}\bar{m}
\end{equation}

By direct differentiation one finds

\begin{equation}
DG_{R} = 0,~~~~~~G_{L} \overleftarrow{D} = 0           \label{207}
\end{equation}

Each term of the series $F_{R}$ satisfies the analytic equation. Thus

\begin{equation}
D g_{n}^{R}(x)=0,~~~~~~g_{n}^{R}= -\frac{\mid p\mid^{2}}{4} \Box (p^{-1} x)^{n}
\end{equation}

Putting

\begin{equation}
p^{-1}x =u,~~~~~~\frac{u}{\mid u\mid }= \xi
\end{equation}
we find

\begin{eqnarray}
g_{n}^{R} &=& -\frac{\mid p\mid^{2}}{4} \Box u^{n}= (\bar{u})^{n-2}~\frac{d}{d\xi}(1+\xi+\cdots+\xi^{n-1}) \nonumber \\
&=& \bar{u}^{n-2}~\frac{d}{d\xi}(\frac{1-\xi^{n}}{1-\xi})
\end{eqnarray}
for $n>1$. We have

\begin{equation}
g_{0}=g_{1}=0
\end{equation}

For negative $n$ we have

\begin{equation}
g_{-n}^{R}= (\bar{u})^{-n-2}~\frac{d}{d\xi} (1+\xi^{-1}+\xi^{-2} + \cdots+\xi^{-n})
\label{212}
\end{equation}
We also have the orthogonality property

\begin{eqnarray}
(g_{n}^{L},g_{m}^{R}) &=& \oint \bar{g}_{n}^{L}~d\Sigma~g_{m}^{R} \nonumber \\ &=& \int
d^{4}x~(\bar{g}_{n}^{L}Dg_{m}^{R} + \bar{g}_{n}^{L} \overleftarrow{D} g_{m}^{R})=0
\end{eqnarray}
for $n\neq m$. Here

\begin{equation}
d\Sigma = d\Sigma_{0} - i \vec{\sigma} \cdot d\vec{\Sigma}
\end{equation}
where $d\Sigma_{\mu}$ is the surface element. The $3$-dimensional surface integral can be
taken over $S^{3}$.

In this way a solution $f$ of the massless Dirac equation

\begin{equation}
Df=0
\end{equation}
can be expanded in the basis functions $g_{n}^{R}$.
Now a solution of the Dirac equation is also a solution of the equation

\begin{equation}
\bar{x} D f =0                       \label{216}
\end{equation}

We can write

\begin{equation}
\bar{x}D=x^{\mu} \partial_{\mu} - \vec{\sigma} \cdot \vec{L} =\Delta-\vec{\sigma} \cdot
\vec{L},
\end{equation}
where

\begin{equation}
\vec{L}=-i(\vec{x} \times \vec{\nabla} + \vec{x} \partial_{0} -x_{0} \vec{\nabla})
\end{equation}
is the self dual part of the angular momentum tensor associated with the 
$SU(2)_{L}$ subgroup of $O(4)$.  The operator $\Delta$ is the dilatation.  Hence $f$ satisfies

\begin{equation}
\Delta f(x)=\vec{\sigma} \cdot \vec{L} ~f(x)
\end{equation}
If $f^{(j)}(x)$ is homogeneous of degree $j$, we have

\begin{equation}
\Delta f^{(j)}(x)f(x)= j~f^{(j)}(x)
\end{equation}

Iterating and using the commutation relation

\begin{equation}
\vec{L} \times \vec{L}=i~\vec{L}
\end{equation}
we find

\begin{equation}
(\vec{\sigma} \cdot \vec{L})^{2} = \vec{L} \cdot \vec{L}-\vec{\sigma} \cdot \vec{L}
\end{equation}
so that

\begin{equation}
\vec{L} \cdot \vec{L} ~f^{(j)}(x) = j(j+1)~f^{(j)}(x)
\end{equation}

Hence $f^{(j)}(x)$ is proportional to a Wigner $D_{mm'}^{j}$ function on $S^{3}$ (or
$SU(2)$).  The formulas (\ref{207})-(\ref{212}) give the relation between the $S^{3}$ functions
and the basis functions of Fueter analytic mappings.
     
To find the relation with the harmonic analyticity of Ogievetsky and his collaborators$^{\cite{c2}}$ 
let us rewrite Eq.(\ref{216}) in $2\times 2$ matrix form. We have

\begin{equation}
x=\left(
\begin{array}{cc}
\bar{u}_{2}  & u_{1} \\
-\bar{u}_{1}  & u_{2}
\end{array}
\right),~~~~~~\bar{x} = \left(
\begin{array}{cc}
u_{2}  & -u_{1} \\
\bar{u}_{1}  & \bar{u}_{2}
\end{array}
\right)
\end{equation}
where

\begin{equation}
u_{1}=-(x^{2} + i x^{1}),~~~~~~u_{2}=x^{0} +i x^{3}
\end{equation}
With these new coordinates $D$ takes the form

\begin{equation}
\partial_{0}-i~\vec{\sigma} \cdot \vec{\nabla}=D= \left(
\begin{array}{cc}
\partial / \partial u_{2}  & \partial /\partial \bar{u}_{1} \\
-\partial / \partial u_{1}  & \partial / \partial \bar{u}_{2}
\end{array}
\right),
\end{equation}
\vskip .5cm
$ i\bar{x}D =  \Delta +\vec{\sigma} \cdot \vec{L}= $

\begin{equation}
= i \left(
\begin{array}{cc}
u_{1} \partial / \partial u_{1}+ u_{2} \partial / \partial u_{2} & -u_{2} \partial /\partial
\bar{u}_{1}- u_{1} \partial /\partial \bar{u}_{2} \\
\bar{u}_{1} \partial / \partial u_{2}-\bar{u}_{2} \partial / \partial u_{1}  & \bar{u}_{1} \partial
/ \partial \bar{u}_{1}+\bar{u}_{2} \partial / \partial \bar{u}_{2}
\end{array}
\right)
\end{equation}
We can rewrite this as

\begin{equation}
i\bar{x}D=i \left(
\begin{array}{cc}
D^{+-} &-D^{++} \\
D^{--} & D^{-+}
\end{array}
\right)
\end{equation}

Let 

\begin{equation}
f= \left(
\begin{array}{cc}
f_{1} &-\bar{f}_{2} \\
f_{2} & \bar{f}_{1}
\end{array}
\right)
\end{equation}
We find

\begin{equation}
D^{++} =L_{2} + iL_{1}
\end{equation}
or if 

\begin{equation}
L_{\mu \nu}=i(x_{\mu} \partial_{\nu} - x_{\nu} \partial_{\mu})
\end{equation}

\begin{eqnarray}
D^{++} &=& u_{1} \partial / \partial \bar{u}_{2}- u_{2} \partial / \partial \bar{u}_{1}  \nonumber \\
&=& L_{20}+L_{31} + i~(L_{23}+L_{10})
\end{eqnarray}
an operator associated with the self-dual angular momentum generating left rotations.  The
$U(1)$ subgroup is generated by

\begin{equation}
L_{3} = \frac{i}{2}(D^{+-}-D^{-+})
\end{equation}
whereas the dilatations giving the degree of homogeneity is given by

\begin{equation}
\Delta = \frac{i}{2}(D^{+-} + D^{-+})
\end{equation}

On $S^{2}$ for a given representation $L_{3}$ and $\Delta$ have given eigenvalues so that
the self-dual analyticity equation (\ref{216}) reduces to an equation in $D^{++}$.  This operator
and the variable $x_{+}$ defined in Eq.(\ref{1.43}) are the cornerstones of harmonic analyticity
which involve functions of $x_{+}$ and $u$, where $u$ represents a spinor or a unit quaternion. 
If $u$ is defined up to a phase it represents a point on $S^{2}$ and functions of $x_{+}$ and
$u$ are equivalent to functions of a twistor or functions of two quaternions. It follows that
twistor analyticity, harmonic analyticity and quaternionic Fueter analyticity are all related.
     
The operator $\bar{x}D$ represents the dilatation and the left rotation subgroups of the
conformal group. Similarly

\begin{equation}
Dx=x_{0}D-i\vec{x}\cdot D \vec{\sigma}
\end{equation}
represents the combination of dilatations and right rotations.  $D$ represents the translations.  In
a special conformal transformation we have from Eq.(\ref{53})

\begin{equation}
\delta_{c}x=c_{\mu} x {\bar{e}}^{\mu} x,~~~~~~\delta_{c} x^{\nu}=
c_{\mu}~Sc(\bar{e}^{\nu} x \bar{e}^{\mu} x)
\end{equation}
where

\begin{equation}
e^{0}=\bar{e}^{0} = 1, ~~~~~e^{n}=-\bar{e}^{n}=-i \sigma^{n}
\end{equation}
Hence for a function $\phi(x)$ we have

\begin{equation}
\delta_{c} \phi(x) = \phi(x+\delta_{c}x)-\phi(x) = c_{\mu}~Sc(x\bar{e}^{\mu} x
\bar{D})\phi(x),
\end{equation}

This equation gives for the generators of special conformal transformation the operators

\begin{equation}
S^{\mu} = Sc(x \bar{e}^{\mu} x \bar{D})
\end{equation}
which in quaternion form read

\begin{equation}
S=e_{\mu} S^{\mu} = x \bar{D} x = x \bar{e}^{\mu} x \partial_{\mu}.
\end{equation}

Thus, the $2\times 2$ quaternionic matrix operator for $SL(2,Q)$ takes the form

\begin{equation}
{\cal{L}}= \left(
\begin{array}{cc}
D\bar{x} & D \\
\bar{x} D \bar{x} & -\bar{x} D
\end{array}
\right)
\end{equation}

\section{Various forms of quaternionic differential operators}

Consider the quaternion

\begin{equation}
U=e^{\mu} ~\xi_{\mu} = \xi_{0}-i\vec{\sigma} \cdot
\vec{\xi},~~~~(\bar{U}=\xi_{0}+i\vec{\sigma} \cdot \vec{\xi})      \label{6.1}
\end{equation}

Written as a $2\times 2$ matrix it has the form

\begin{equation}
U= \left(
\begin{array}{cc}
\xi_{0}-i\xi_{3}  & -i\xi_{1}-\xi_{2}  \\
i\xi_{1}+\xi_{2}  &  \xi_{0}+i\xi_{3}
\end{array}
\right) = \rho~u
\end{equation}
where

\begin{equation}
\rho = (U \bar{U})^{\frac{1}{2}} =
{\sqrt{\xi_{0}^{2}+\xi_{1}^{2}+\xi_{2}^{2}+\xi_{3}^{2}}}= {\sqrt{{\rm Det}~U}}                  
\label{6.3}
\end{equation}
and

\begin{equation}
u \bar{u} = u u^{\dag} = 1
\end{equation}
so that u is unitary and has the form

\begin{equation}
u= \frac{1}{\rho} \left(
\begin{array}{cc}
\bar{u}_{2} & u_{1} \\
-\bar{u}_{1} & u_{2}
\end{array}
\right)
\end{equation}
with

\begin{equation}
u_{1} = -(\xi_{2} +i \xi_{1}),~~~~~~u_{2}=\xi_{0}+i\xi_{3}         \label{6.6}
\end{equation}

The unitary matrix $u$ is an element of $SU(2)$. Its subgroup $U(1)$ can be chosen as the
diagonal subgroup

\begin{equation}
h(\psi) = exp(-i\sigma_{3} \psi)
\end{equation}
Thus

\begin{equation}
u=Z(z)~h(\psi)
\end{equation}
where the complex $z$ parametrizes the coset $Z \in SU(2)/U(1)$ which has the form

\begin{equation}
Z(z) = (1+z \bar{z})^{-\frac{1}{2}} \left(
\begin{array}{cc}
1 & z \\
-\bar{z} & 1
\end{array}
\right)
\end{equation}
and represents the sphere $S^{2}= S^{3}/S^{1}$.

Thus, instead of the real $4$-coordinate $\xi_{\mu}$ we can use the two complex coordinates
$u_{1}$, $u_{2}$ or the coordinates $\rho$, $z$ and $\psi$ with $\rho$ real and positive, $\psi$
real and in the interval $(0,\pi)$ and $z$ complex.

The relations are given by Eqs.(\ref{6.6}) and (\ref{6.3}) and

\begin{equation}
z = - \frac{\xi_{2} + i \xi_{1}}{\xi_{0} +i \xi_{3}} = \frac{u_{1}}{u_{2}}
\end{equation}
\begin{equation}
e^{2i \psi} = \frac{\xi_{0} + i \xi_{3}}{\xi_{0} -i \xi_{3}}=\frac{u_{2}}{\bar{u}_{2}}
\end{equation}
or

\begin{equation}
\psi = \frac{1}{2i} (ln \frac{\xi_{0} + i \xi_{3}}{\xi_{0} -i \xi_{3}})
\end{equation}

Also

\begin{equation}
u_{1} = -(\xi_{2} + i \xi_{1}) = \rho~\frac{z e^{i\psi}}{\sqrt{1+z\bar{z}}}  \label{6.12a}
\end{equation}

\begin{equation}
u_{2} = \xi_{0} + i \xi_{3} = \rho~\frac{ e^{i\psi}}{\sqrt{1+z\bar{z}}}  \label{6.12b}
\end{equation}

\begin{equation}
\rho^{2} = u_{1} \bar{u}_{1}+u_{2} \bar{u}_{2} = \xi_{\mu} \xi^{\mu}  \label{6.12c}
\end{equation}

We now construct the quaternionic differential operators $D$ and $\bar{D}$ given by

\begin{equation}
D= e_{\mu} ~\frac{\partial}{\partial \xi_{\mu}} = \frac{\partial}{\partial \xi_{0}} -
i\vec{\sigma} \cdot \frac{\partial}{\partial \vec{\xi}}
\end{equation}

\begin{equation}
\bar{D}= \frac{\partial}{\partial
\xi_{0}} + i\vec{\sigma} \cdot \frac{\partial}{\partial \vec{\xi}}
\end{equation}
$D$ has the matrix form

\begin{equation}
D= \left(
\begin{array}{cc}
\partial_{0} -i \partial_{3} & -i \partial_{1} - \partial_{2}  \\
-i \partial_{1} + \partial_{2}  & \partial_{0} +i \partial_{3}
\end{array}
\right)
\end{equation}
where

\begin{equation}
\partial_{\mu} = \frac{\partial}{\partial \xi^{\mu}} 
\end{equation}
In terms of $u_{1}$, $u_{2}$ we have

\begin{equation}
D=2 \left(
\begin{array}{cc}
\partial / \partial u_{2} & \partial / \partial \bar{u}_{1}  \\
-\partial / \partial u_{1} & \partial / \partial \bar{u}_{2}
\end{array}
\right)
\end{equation}

\begin{equation}
\bar{D}=2 \left(
\begin{array}{cc}
\partial /\partial \bar{u}_{2} & -\partial / \partial \bar{u}_{1}  \\
\partial / \partial u_{1} & \partial / \partial u_{2}
\end{array}
\right)
\end{equation}

We also have

\begin{equation}
\frac{\partial}{\partial \bar{u}_{1}}= e^{i\psi} (\frac{1}{2} \frac{z}{\sqrt{1+z\bar{z}}}
\frac{\partial}{\partial \rho}+\frac{1}{\rho} \sqrt{1+z\bar{z}} \frac{\partial}{\partial \bar{z}}),  
              \label{6.16a}
\end{equation}

\begin{eqnarray}
\frac{\partial}{\partial \bar{u}_{2}} &=& e^{i\psi} (\frac{1}{2} \frac{z}{\sqrt{1+z\bar{z}}}
\frac{\partial}{\partial \rho}-\frac{\bar{z}}{\rho} \sqrt{1+z\bar{z}} \frac{\partial}{\partial
\bar{z}}  \nonumber  \\
 &~~~~~-& \frac{1}{2i} \frac{1}{\rho} \sqrt{1+z\bar{z}} \frac{\partial}{\partial \psi})                
\label{6.16b}
\end{eqnarray}

Consider the dilatation operator

\begin{equation}
\Delta = \xi^{\mu} \frac{\partial}{\partial \xi^{\mu}} = \frac{1}{2}
(u_{1}\frac{\partial}{\partial u_{1}}+u_{2}\frac{\partial}{\partial u_{2}}+
\bar{u}_{1}\frac{\partial}{\partial \bar{u}_{1}} + \bar{u}_{2}\frac{\partial}{\partial
\bar{u}_{2}})
\end{equation}
and the covariant angular momentum operator

\begin{equation}
L_{\mu \nu} = i(\xi_{\mu} \partial_{\nu} - \xi_{\nu} \partial_{\mu})
\end{equation}

Its self dual part is

\begin{equation}
\ell_{\mu \nu}=\frac{1}{2}  (L_{\mu \nu} + \tilde{L}_{\mu \nu})
\end{equation}
where

\begin{equation}
\tilde{L}_{\mu \nu} = \frac{1}{2} \epsilon_{\mu \nu \alpha \beta} L^{\alpha \beta}
\end{equation}

Let

\begin{equation}
\ell_{1} = \ell_{23} = \ell_{01},~~~~\ell_{2} = \ell_{31} = \ell_{02},~~~~\ell_{3} = \ell_{12}
= \ell_{03}
\end{equation}
We find

\begin{equation}
{\rm Sc}~\bar{U}D = \frac{1}{2} Tr~\bar{U} D = \Delta
\end{equation}

\begin{equation}
{\rm Vec}~\bar{U}D = \bar{U}D-\frac{1}{2} I Tr(\bar{U}D)= -i \vec{\sigma} \cdot \vec{\ell}
\end{equation}
where $\bar{U}$ is given by Eq.(\ref{6.1}). Let
\begin{equation}
\Omega = \frac{1}{2} \bar{U}D= \frac{1}{2} (\Delta-i\vec{\sigma}\cdot
\vec{\ell})=\frac{1}{2}
\left(
\begin{array}{cc}
D^{+-} & -D^{++} \\
D^{--} & D^{-+}
\end{array}
\right)
\end{equation}
We find

\begin{equation}
D^{+-} = \Delta - i \ell_{3} = u_{1}\frac{\partial}{\partial u_{1}} +u_{2}\frac{\partial}{\partial u_{2}}
\end{equation}

\begin{equation}
D^{-+} = \Delta + i \ell_{3} = \bar{u}_{1}\frac{\partial}{\partial \bar{u}_{1}}
+\bar{u}_{2}\frac{\partial}{\partial \bar{u}_{2}}
\end{equation}

\begin{equation}
D^{--} =  \bar{u}_{1}\frac{\partial}{\partial u_{2}} -\bar{u}_{2}\frac{\partial}{\partial
u_{1}}
\end{equation}

Using Eqs.(\ref{6.12a}) to (\ref{6.12c}), (\ref{6.16a}) and (\ref{6.16b}) we find

\begin{equation}
D^{++}=e^{2i\psi}[-(1+z\bar{z})\frac{\partial}{\partial \bar{z}}+\frac{i}{2} z
\frac{\partial}{\partial \psi}]
\end{equation}

\begin{equation}
D^{-+}= \rho \frac{\partial}{\partial \rho} +\frac{i}{2} \frac{\partial}{\partial \psi}
\end{equation}

\begin{equation}
D^{+-}= \rho \frac{\partial}{\partial \rho} -\frac{i}{2} \frac{\partial}{\partial \psi} =
(D^{-+})^{\ast}
\end{equation}

\begin{equation}
D^{--} = -(D^{++})^{\ast}
\end{equation}
or

\begin{equation}
\Delta=\xi^{\mu} \partial_{\mu} = \rho \frac{\partial}{\partial \rho}
\end{equation}

\begin{equation}
\ell_{3} = \frac{i}{2} \frac{\partial}{\partial \psi}
\end{equation}

\begin{equation}
\ell_{1}-i\ell_{2} = i e^{2i\psi}[-(1+z\bar{z})\frac{\partial}{\partial \bar{z}} +
\frac{i}{2} z \frac{\partial}{\partial \psi}]
\end{equation}
Note that $\Delta$ depends only on $\rho$ and $\vec{\ell}$ does not depend on $\rho$, while
$\ell_{3}$ depends only on $\psi$ as it should be.

\section{Factorization of the Energy-Momentum tensor as the analog of the 
Sommerfield-Sugawara factorization in $D=2$. Generalizations.}

In a large class of $D=2$ chiral field theories we are led to the 
Sommerfield-Sugawara$^{\cite{c9}}$,$^{\cite{c10}}$  form of the traceless energy-momentum tensor

\begin{equation}
T_{\alpha \beta} = J_{\alpha}^{i}(x)~J_{\beta i}(x)-\frac{1}{2} \delta_{\alpha \beta}~J^{i
\gamma}~J_{i \gamma},
\end{equation}
where $j_{\alpha}^{i}$ and their duals $\tilde{J}_{\alpha}^{i}$ are conserved currents of the
theory

\begin{equation}
\partial_{\alpha}~J^{\alpha i}=0,~~~~~~~\partial_{\alpha}~\tilde{J}^{\alpha
i}=\epsilon^{\alpha \beta} \partial_{\alpha}~J_{\beta}^{i}=0
\end{equation}
Then

\begin{equation}
\partial_{\alpha}~T^{\alpha}_{\beta} = (\partial_{\alpha}~J^{i \alpha})~J_{i \beta}+
J^{\alpha i}~\partial_{\alpha}~J^{i}_{\beta} - J^{\alpha i}~\partial_{\beta}~J_{\alpha i}=0
\end{equation}

It is known that the moments of $J^{\alpha i}$ generate a Kac-Moody algebra, whereas those of
$T^{\alpha}_{\beta}$ satisfy a Virasoro algebra$^{\cite{c11}}$.
     
In conformally invariant $D=4$ theories the energy momentum tensor $T_{\mu \nu}$ is
traceless and symmetric.  It can be represented by a traceless symmetric $4\times 4$ matrix $T$
associated with the representation $(1,1)$ of $O(4)$.

Hence it should be possible to factorize it into $(1,0)$ self-dual fields $F_{\alpha \beta}^{i}$
and $(0,1)$ anti self-dual fields $G_{\alpha \beta}^{i}$ so that

\begin{equation}
T_{\alpha \beta} = F_{\alpha \rho}^{i}(x)~G_{\beta i}^{\rho}(x)
\end{equation}

Now quaternions have two real $4\times 4$ matrix representations associated respectively with
left and right quaternionic multiplications.  Since these two operations commute so do the matrix
representations.  For purely vectorial quaternions one corresponds to a self-dual antisymmetric
matrix, while the other is anti self dual and antisymmetric.  They have the forms

\begin{equation}
F=\vec{e}_{L} \cdot \vec{f}=\left(
\begin{array}{cccc}
0     & -f_{3} & f_{2} & f_{1} \\
f_{3} &   0    &  -f_{1} & f_{2} \\
-f_{2} & -f_{1} &  0     &  f_{3} \\
-f_{1} &  -f_{2} & -f_{3} &  0 
\end{array}
\right)= \vec{e} \cdot \vec{f}
\end{equation}
and

\begin{equation}
G' =\vec{e}_{R} \cdot \vec{g}=\left(
\begin{array}{cccc}
0      & -g_{3} &  g_{2} &  -g_{1} \\
g_{3} &   0    &  -g_{1} & -g_{2} \\
-g_{2} & g_{1} &  0     &  -g_{3} \\
g_{1} &  g_{2} & g_{3} &  0 
\end{array}
\right) = \vec{e}^{~'} \cdot \vec{g}
\end{equation}
and we have

\begin{equation}
[F,G']=0,~~~~~F^{2}=-\vec{f} \cdot \vec{f}~I,~~~~~G'^{2}=-\vec{g} \cdot \vec{g}~I
\end{equation}
In the case of one internal degree of freedom we have

\begin{equation}
T=F(x)~G'(x)=G'(x)~F(x)
\end{equation}
or

$$
T = T^{T}=~~~~~~~~~~~~~~~~~~~~~~~~~~~~~~~~~~~~~~~~~~~~~~~~~~~~~~~~~~~~~~~~~~~~
$$

\begin{equation}
= \left(
\begin{array}{cccc}
2f_{1} g_{1}- \vec{f} \cdot \vec{g} & f_{1}g_{2}+f_{2}g_{1} & f_{3}g_{1}+f_{1}g_{3} &
-(\vec{f} \times \vec{g})_{1} \\
f_{1}g_{2}+f_{2}g_{1} & 2f_{2} g_{2}- \vec{f} \cdot \vec{g} & f_{2}g_{3}+f_{3}g_{2} &
-(\vec{f} \times \vec{g})_{2} \\
f_{3}g_{1}+f_{1}g_{3} & f_{2}g_{3}+f_{3}g_{2} & 2f_{3} g_{3}- \vec{f} \cdot \vec{g} &
-(\vec{f} \times \vec{g})_{3} \\
-(\vec{f} \times \vec{g})_{1} & -(\vec{f} \times \vec{g})_{2} & -(\vec{f} \times \vec{g})_{3}
& \vec{f} \cdot \vec{g}
\end{array}
\right)~~~
\end{equation}
T is symmetric and traceless. On putting

\begin{equation}
\vec{f}=\frac{1}{2}(\vec{E}+\vec{B}),~~~~~~\vec{g}=\frac{1}{2}(\vec{E}-\vec{B})
\end{equation}
one obtains the familiar form of Maxwell's tensor in euclidean space-time.  When $\vec{f}$ and
$\vec{g}$ are labeled by an internal symmetry index $i$ they can be interpreted as the self dual
and anti self dual part of an anti-symmetric Yang-Mills field tensor in euclidean space-time. 
Then

\begin{equation}
\vec{f}^{i}=0,~~~~~~~~(F^{i}=0)
\end{equation}
becomes the self-duality equation.

Note that $\vec{f}^{i}$ are not currents but self-dual fields.  Currents $J_{\mu}^{i}$ are
represented by $4\times 4$ quaternions with trace $4 J_{0}^{i}$ which transform like vectors
under $O(4)$.  In the Maxwell case we have

\begin{equation}
J=F~\overleftarrow{\bar{D}}=D~G' =\frac{1}{2}(D~G'+F~\overleftarrow{\bar{D}})
\end{equation}
The homogeneous set of Maxwell's equations reads

\begin{equation}
D~G'-F~\overleftarrow{\bar{D}}=0
\end{equation}
The current is given by

\begin{equation}
J=J_{0}+\vec{\sigma}_{R} \cdot \vec{J}
\end{equation}
and it transforms like the $(\frac{1}{2},\frac{1}{2})$ representation of $O(4)$. 

The conservation of $J$ is given by

\begin{equation}
{\rm Sc}~\bar{D} G' = {\rm Sc}~\Box G' =0
\end{equation}
since $G'$ is traceless.  In the anti self-dual case $G'=0$ so that $J=0$ as expected.  The same
result follows from  self duality.

Thus, the decomposition of $T$ in terms of currents is

\begin{equation}
T=(D^{-1}J)(J \overleftarrow{\bar{D}}^{-1})
\end{equation}
where $D^{-1}$ is the Dirac Green's function in the massless case.

When $J=0$ we can obtain $F$ and $G'$ as quaternionic analytic functions, since for example

\begin{equation}
D~G' =0
\end{equation}
 
Then 

\begin{equation}
G' =\Box \sum_{n} c_{n} (p^{-1} x)^{n}
\end{equation}
which is also a solution of 

\begin{equation}
\bar{x}~D~G' = 0
\end{equation}
associated with harmonic analyticity as we have seen earlier.
     
The energy momentum tensor $T$ is not the only tensor that can be constructed from $F$ and
$G'$.  In fact we can obtain a scalar $(0,0)$, a $s=2$ for $SU_{L}(2)$ [$(2,0)$ of $O(4)$]
represented by self-dual Weyl tensor $W_{L}$ and $s=2$ for $SU_{R}(2)$ [$(0,2)$ of $O(4)$]
represented by an anti self dual Weyl tensor $W_{R}$.  The $(2,0)$ field is represented by a
traceless $3\times 3$ symmetric matrix. It can be combined with the $(0,0)$ scalar $\ell$ into a
$4\times 4$ matrix $L$ while a scalar $r$ is combined with the $(0,2)$ into another $4\times 4$
matrix $R$. Introducing the diagonal matrix $\Delta$

\begin{equation}
\Delta = \left(
\begin{array}{cccc}
\frac{1}{3} & 0 & 0 & 0 \\
0 & \frac{1}{3} & 0 & 0 \\
0 & 0 & \frac{1}{3} & 0 \\
0 & 0 & 0 & -\frac{2}{3}
\end{array}
\right)
\end{equation}
We find

\begin{eqnarray}
L=L^{T}= \left(
\begin{array}{cc}
W_{L} & 0 \\
0 & \ell
\end{array}
\right) = F~\Delta~F=~~~~~~~~~~~~~~~~~~~~~~& &     \nonumber \\
=\left(
\begin{array}{cccc}
f_{1}^{2}-\frac{1}{3} \vec{f}\cdot \vec{f} & f_{1} f_{2} & f_{3} f_{1} & 0 \\
f_{1} f_{2} & f_{2}^{2}-\frac{1}{3} \vec{f}\cdot \vec{f} & f_{2} f_{3} & 0 \\
f_{3} f_{1} & f_{2} f_{3} & f_{3}^{2}-\frac{1}{3} \vec{f}\cdot \vec{f} & 0 \\
0 & 0 & 0 & -\frac{1}{3} \vec{f}\cdot \vec{f}
\end{array}
\right)
\end{eqnarray}
and similarly for $R$

\begin{equation}
R=\left(
\begin{array}{cc}
W_{R} & 0 \\
0 & r
\end{array}
\right) = G'~\Delta~G'
\end{equation}

We have

\begin{equation}
-3 \ell =\vec{f} \cdot \vec{f}= \frac{1}{4} (\vec{E} +\vec{B})^{2} = \frac{1}{4} (
\vec{E}^{2} +\vec{B}^{2}) + \frac{1}{2} \vec{E} \cdot \vec{B}
\end{equation}

\begin{equation}
-3 r =\vec{g} \cdot \vec{g}= \frac{1}{4} (\vec{E} -\vec{B})^{2} = \frac{1}{4} ( \vec{E}^{2}
+\vec{B}^{2}) - \frac{1}{2} \vec{E} \cdot \vec{B}
\end{equation}
giving the two invariants of Maxwell's theory in the euclidean case.

Here we note that the curvature tensor in a Riemannian euclidean tensor can be decomposed into
$5$ pieces transforming under $O(4)$ as follows$^{\cite{c12}}$
\begin{equation}
R =(0,0),~~~~~~~{\rm scalar~~ curvature}
\end{equation}

\begin{equation}
R_{\mu \nu}-\frac{1}{4} g_{\mu \nu} R =(1,1),~~~~~~~{\rm traceless ~~Ricci~~ tensor}
\end{equation}

\begin{equation}
C^{L}_{\alpha \beta \mu \nu} =(2,0),~~~~~~~{\rm self~~ dual~~ Weyl~~ tensor}
\end{equation}

\begin{equation}
C^{R}_{\alpha \beta \mu \nu} =(2,0),~~~~~~~ {\rm anti~~ self~~ dual~~ Weyl~~ tensor}
\end{equation}

In the case these $20$ functions can be expressed quadratically in terms of the six components
$\vec{f}$ and $\vec{g}$ of an antisymmetrical tensor $\phi_{\mu \nu}$, its 
pieces can be factorized into $F~G'$, $F~\Delta~F$ and $G'~\Delta~G'$.  If $G'=0 $ (self dual
$\phi$) then the Ricci tensor as well as $W_{R}$ vanish and we have a half-flat space.  All 
the curvature can be expressed in terms of $F$ which has the quaternion analytic form

\begin{equation}
F(x)= \Box \sum_{n} c_{n} (xp^{-1})^{n}
\end{equation}

When $F$ is meromorphic we have a gravitational instanton solution.  These results generalize
to the Yang-Mills case with $F~\Delta~F \rightarrow F^{i}~\Delta~F_{i}$, etc.

This shows that the factorization of the self-dual Weyl tensor is a better analog to the
Sommerfield-Sugawara factorization of $T$ in $D=2$, since both factors $F$ can be analytic,
whereas $T$ in $D=4$ requires one analytic factor $(1,0)$ and one anti analytic factor $(0,1)$.

For an operator product expansion of $L$ for example we would have

\begin{equation}
L(x)L(x')=\left(
\begin{array}{cc}
W_{L}(x)~W_{L}(x') & 0 \\
0 & \ell (x)~\ell (x')
\end{array}
\right)
\end{equation}
$W_{L}(x)~W_{L}(x')$ is a $3\times 3$ matrix that decomposes into a multiple of unity
($s=0$), an antisymmetrical matrix ($s=1$) and a symmetric traceless matrix.  Thus we can write 

\begin{eqnarray}
W_{L}(x)~W_{L}(x')&=&  K(x-x')~I + u(x-x')~F(x) \nonumber \\
   &~~~~~~& + v(x-x')~W_{L}(x)
\end{eqnarray}
with singular $c$-number coefficients that can be determined from dimensional arguments.  The
product $T(x)~T(x')$ is more complicated and bears less resemblance to the $D=2$ case.  We
conclude that the analog of $T(z)$ in $D=2$ is $W_{L}(x)$ which has the simplest operator 
product expansion.

To test its true usefulness and power, this quaternionic formalism must still await further 
progress in the representation theory of $D=4$ diffeomorphism and current algebras and the 
discovery of $D=4$ counterpart of the Wess-Zumino-Novikov-Witten model.

Understanding of geometry of supersymmetry through harmonic analyticity and harmonic superspace led into wide variety of applications in physics and geometry$^{\cite{howe}}$. It has potentially interesting application to some on-shell supermultiplets which arise naturally in string theory in the context of Maldacena conjecture$^{\cite{maldacena}}$ (which relates supergravity theories on anti-de Sitter backgrounds to conformal field theories on the boundary, i.e. super-Minkowski space). 

The search for exceptional structures associated with eleven dimensions has lead us into description of space-time by means of exceptional groups$^{\cite{ramzi}}$. A quest toward a unified theory has taken us through a progression from particle world lines of quantum field theory (${\cal{R}}$ analyticity) to string worldsheet in order to incorporate gravity (${\cal{C}}$ analyticity) into $M$-brane worldvolumes in order to realize duality (${\cal{H}}$ analyticity). Next step would be the search for octonionic analyticity ($ \Omega$ analyticity). $M$-theory might turn out to be supersymmetric in eleven dimensions. 

Octonionic planes may possibly provide the geometrical foundation for the existence of internal symmetries like color and flavor. These octonionic geometries allow us to construct new finite Hilbert spaces with unique properties. The non-Desargues'ian geometric property makes them non-embeddeble in higher spaces, hence essentially finite. It also leads to peculiarities in the superposition principle in the color sector of the Hilbert space. This theory of the charge space, if correct, suggests a new geometric picture for the substructure of the material world, that of an octonionic geometry attached at each point of Einstein's Riemannian manifold for space time with local symmetry that leaves the properties of charge space invariant$^{\cite{ramond}, \cite{sultan}}$.  Extension of the formalism we presented to octonionic structures may help to understand geometry, duality and non perturbative physics in a deeper way.

Four dimensional conformal field theory can also be formulated on Kulkarni$^{\cite{kulkarni}}$ $4$-folds leading to a formalism that parallels that of $2$D conformal field theory on Riemann surfaces.$^{\cite{zucchini}}$ In this framework the notion of Fueter analyticity, the quaternionic analogue of complex analyticity plays an essential role. For other approaches where symmetry of some field equations were investigated, see for example paper by Kruglov.$^{\cite{kruglov}}$
\newpage 
\section*{Appendix: Use of Coherent States in $SU(2)$}

Let 

\begin{equation}
a= \pmatrix{ a_1 \cr
a_2 \cr}~,~~~~a^\dag = \pmatrix{ a_1^\dag & a_2^\dag \cr}
\end{equation}
and

\begin{equation}
\xi= \pmatrix{ \xi_1 \cr \xi_2 \cr}~,~~~~\xi^\dag = \pmatrix{\xi_1^* & \xi_2^* \cr}
\end{equation}
with

\begin{equation}
[a_i,a_j^\dag]=\delta_{ij} 
\end{equation}
and defining the vacuum state $c=\mid 0>$, we have

\begin{equation}
a~\mid 0>=0~,~~~~<0\mid ~a^\dag=0
\end{equation}

We can construct a superposition state (coherent state) out of the vacuum state

\begin{equation}
\mid \xi>= e^{a^\dag \xi} ~\mid 0>~, ~~~~<\xi\mid =<0\mid ~e^{\xi^\dag a}
\end{equation}
that has the following properties:

(1.)

\begin{equation}
<0\mid \xi> =<\xi\mid 0> =<0\mid 0>=1
\end{equation}

(2.)
\begin{eqnarray}
\mid \xi>&=& e^{a_1^\dag \xi_1}~e^{a_2^\dag \xi_2}~ \mid 0> \\ 
& & = \sum_{j=0}^\infty~\sum_{m=-j}^j~ \frac{a_1^{\dag j+m}}{\sqrt{(j+m)!}}~ 
\frac{a_2^{\dag j-m}}{\sqrt{(j-m)!}}~\frac{\xi_1^{j+m}}{\sqrt{(j+m)!}}~\frac{\xi_2^{j-m}}{\sqrt{(j-m)!}}   
\end{eqnarray}
or

\begin{equation}
\mid \xi>= \sum_{j=0}^\infty~\sum_{m=-j}^j ~\frac{\xi_1^{j+m}}{\sqrt{(j+m)!}}~\frac{\xi_2^{j-m}}{\sqrt{(j-m)!}}~\mid j,m>   
\end{equation}
with

\begin{equation}
<\xi\mid = \sum_{j=0}^\infty~\sum_{m=-j}^j ~<j,m\mid  ~\frac{\xi_1^{*~j+m}}{\sqrt{(j+m)!}}~\frac{\xi_2^{*~j-m}}{\sqrt{(j-m)!}}
\end{equation}

(3.)

\begin{equation}
e^{-a^\dag \xi}~a_1~e^{a^\dag \xi} = e^{-a_1^\dag \xi_1} ~ a_1~e^{a_1^\dag \xi_1} = e^{\xi_1 \frac{\partial}{\partial a_1}}~a_1 = a_1+\xi_1
\end{equation}

\begin{equation}
e^{-a^\dag \xi}~a_2~e^{a^\dag \xi}=a_2+\xi_2 
\end{equation}
so that

\begin{equation}
e^{-a^\dag \xi}~a_i~e^{a^\dag \xi}=a_i +\xi_i 
\end{equation}

(4.)

\begin{equation}
a_i~\mid \xi>= \xi_i ~\mid \xi>
\end{equation}
so that $\mid \xi>$ is an eigenstate of $a_i$. We see that

\begin{eqnarray}
a_i~\mid \xi>&=& a_i~e^{a^\dag \xi} ~\mid 0> = e^{a^\dag \xi} ~e^{-a^\dag \xi}~a_i~ ~e^{a^\dag \xi}~\mid 0>=e^{a^\dag \xi} (a_i+\xi_i)~\mid 0>  \\ \nonumber & & =e^{a^\dag \xi}~\xi_i ~\mid 0>= \xi_i ~\mid \xi>
\end{eqnarray}
and similarly

\begin{equation}
<\xi\mid ~ a_i^\dag = <\xi\mid ~\xi_i^*
\end{equation}

(5.)

\begin{eqnarray}
<\xi\mid j,m>&=&\sum_{j'}~\sum_{m'}~
 ~\frac{\xi_1^{*~j'+m'}}{\sqrt{(j'+m')!}}~\frac{\xi_2^{*~j'-m'}}{\sqrt{(j'-m')!}}~<j',m'\mid j,m> \nonumber \\ & & = \sum_{j'} ~\sum_{m'}~  \frac{\xi_1^{*~j'+m'}}{\sqrt{(j'+m')!}}~\frac{\xi_2^{*~j'-m'}}{\sqrt{(j'-m')!}}~\delta_{j' j} ~\delta_{m' m}
\end{eqnarray}
or

\begin{equation}
<\xi \mid  j,m>=\frac{\xi_1^{*~j+m}}{\sqrt{(j+m)!}}~\frac{\xi_2^{*~j-m}}{\sqrt{(j-m)!}}
\end{equation}
is homogeneous in $\xi_1^*$, $\xi_2^*$ of degree $2j$ so that

\begin{equation}
<\lambda ~\xi\mid  j,m>= \lambda^{* 2j}~<\xi\mid j,m>
\end{equation}
and

\begin{equation}
<j,m\mid \lambda~ \xi>= \lambda^{2j}~<j,m\mid \xi>
\end{equation}

(6.)

\begin{eqnarray}
\frac{1}{\sqrt{(j+m)!}}~\frac{1}{\sqrt{(j-m)!}}~ (\frac{\partial}{\partial \xi_1^*})^{j+m}~(\frac{\partial}{\partial \xi_2^*})^{j-m}~<\xi\mid j,m>= \nonumber 
\end{eqnarray}
\begin{equation}
~~= \frac{1}{(j+m)!}~ \frac{1}{(j-m)!}~ (\frac{\partial}{\partial \xi_1^*})~(\frac{\partial}{\partial \xi_2^*})~\xi_1^{* j+m}~\xi_2^{* j-m}=1
\end{equation}

(7.)

\begin{equation}
a_i^\dag ~\mid \xi>= a_i^\dag~ e^{a_1^\dag \xi_1 + a_2^\dag \xi_2}~\mid 0>= \frac{\partial}{\partial \xi_i}~ e^{a^\dag \xi}~\mid 0>= \frac{\partial}{\partial \xi_i}~\mid \xi>
\end{equation}
Also

\begin{equation}
<\xi\mid ~a_i= \frac{\partial}{\partial \xi_*}~<\xi\mid 
\end{equation}
so that on $\mid \xi>$ we have $a_i \leftrightarrow \xi_i$, $a_i^\dag \leftrightarrow \frac{\partial}{\partial \xi_i} $.

(8.)

\begin{eqnarray}
& & <\xi\mid ~\frac{a_1^{j+m}}{\sqrt{(j+m)!}}~\frac{a_2^{j-m}}{\sqrt{(j-m)!}}~\mid j,m>=1 \nonumber \\
& & ~~~=<j,m\mid j,m>=<0\mid ~\frac{a_1^{j+m}}{\sqrt{(j+m)!}}~\frac{a_2^{j-m}}{\sqrt{(j-m)!}}~\mid j,m>
\end{eqnarray}

\section*{The D functions of $SU(2)$}

Since

\begin{equation}
J_i = a^\dag~\frac{\sigma_i}{2}~a
\end{equation}

\begin{equation}
J_i~\mid 0>=0,~~~~[J_i,J_j]=i~\epsilon_{ijk}~J_k
\end{equation}
and the relations

\begin{equation}
J_1+iJ_2= a_1^\dag~a_2,~~~~J_1-iJ_2= a_2^\dag~a_1,~~~~J_3= \frac{1}{2}~(a_1^\dag a_1 -a_2^\dag a_2)
\end{equation}

\begin{equation}
[J_3, J_1 \pm iJ_2] = J_1 \pm iJ_2~,~~~~~~[J_1 +iJ_2, J_1-iJ_2]=2J_3~,
\end{equation}
we have

\begin{equation}
e^{iJ_3\alpha} ~a^\dag~e^{-iJ_3 \alpha} = a^\dag~ e^{i\sigma_3\frac{\alpha}{2}}
\end{equation}
and

\begin{equation}
e^{i J_2 \beta}~ a^\dag~ e^{-i J_2 \beta} = a^\dag~ e^{i\sigma_2\frac{\beta}{2}}~.
\end{equation}

Now let

\begin{eqnarray}
U(\mbox{\boldmath$\omega$}) ~\mid j,m>&=& e^{iJ_3\alpha}~e^{iJ_2 \beta}~e^{iJ_3 \gamma}~ \mid j,m> \nonumber \\
& &= U~\frac{a_1^{\dag j+m}}{\sqrt{(j+m)!}}~\frac{a_2^{\dag j-m}}{\sqrt{(j-m)!}}~U^{-1}~U~\mid 0>
\end{eqnarray}
with

\begin{equation}
U~\mid 0>=\mid 0>~,~~~~U~a_1^\dag~U^{-1}=a_1^{'\dag}~,~~~~U~a_2^\dag~U^{-1}=a_2^{'\dag}~,
\end{equation}
where

\begin{equation}
a^\dag = \pmatrix{a_1^\dag & a_2^\dag \cr} 
\end{equation}
so that

\begin{eqnarray}
a^{'\dag}~U~a^\dag~U^{-1} &=& e^{iJ_3 \alpha} ~e^{iJ_2\beta}~ e^{iJ_3\gamma}~a^\dag~ e^{-iJ_3\gamma}~ e^{-iJ_2\beta}~e^{-iJ_3\alpha} \nonumber  \\
& & =a^\dag~e^{i\sigma_3 \frac{\alpha}{2}}~ e^{i\sigma_2 \frac{\beta}{2}}~e^{i\sigma_3 \frac{\gamma}{2}} \nonumber  \\
& & =a^\dag~\pmatrix{ \psi_1 & -\psi_2^* \cr
\psi_2 & \psi_1^*} \nonumber  \\
& & = \pmatrix{ a^\dag \psi & a^\dag \hat{\psi}\cr}=\pmatrix{ a_1^{'\dag} & a_2^{'\dag}\cr }
\end{eqnarray}
where 

\begin{equation}
\psi_1= e^{i(\alpha + \gamma)/2}~cos~(\frac{\beta}{2})~,~~~~\psi_2= - e^{i(\gamma-\alpha)/2}~sin(\frac{\beta}{2})
\end{equation}
so that $U(\mbox{\boldmath$\omega$}) =U(\psi)$, and $\psi^\dag \psi=1$.

Also 
\begin{eqnarray}
U(\mbox{\boldmath$\omega$}) \mid j,m'>&=& \frac{a_1^{'\dag j+m'}}{\sqrt{(j+m')!}}~\frac{a_2^{'\dag j-m'}}{\sqrt{(j-m')!}}~\mid 0> \nonumber \\ & & = \frac{(a^\dag \psi)^{j+m'}}{\sqrt{(j+m')!}}~\frac{(a^\dag \hat{\psi})^{j-m'}}{\sqrt{(j-m')!}}~\mid 0>
\end{eqnarray}

Now we can write

\begin{eqnarray}
& &D_{m m'}^j(\mbox{\boldmath$\omega$})= D_{mm'}^j(\psi) = <j, m\mid U(\mbox{\boldmath$\omega$}) \mid j, m'> \nonumber \\ & &
~~~=\frac{(a^\dag \psi)^{j+m'}}{\sqrt{(j+m')!}}~\frac{(a^\dag \hat{\psi})^{j-m'}}{\sqrt{(j-m')!}}~\mid 0> \nonumber  \\ & &
~~~=<0\mid  \frac{a_1^{j+m}}{\sqrt{(j+m)!}}~ \frac{a_2^{j-m}}{\sqrt{(j-m')!}}~\frac{(a^\dag \psi)^{j+m'}}{\sqrt{(j+m')!}}~\frac{(a^\dag \hat{\psi})^{j-m'}}{\sqrt{(j-m')!}}~\mid 0>
\end{eqnarray}

Now to get the expression for $D_{mm'}^j(\psi)$, consider

\begin{equation}
<\xi \mid  U(\psi)\mid j,m'> =<\xi\mid \frac{(a^\dag \psi)^{j+m'}}{\sqrt{(j+m')!}}~\frac{(a^\dag \hat{\psi})^{j-m'}}{\sqrt{(j-m')!}}~\mid 0>
\end{equation}
Using

\begin{equation}
<\xi\mid ~a^\dag=<\xi\mid ~\xi^\dag
\end{equation}
and $<\xi\mid 0>=1$ we get

\begin{equation}
<\xi \mid  U(\psi)\mid j,m'> = \frac{(\xi^\dag \psi)^{j+m'}}{\sqrt{(j+m')!}}~\frac{(\xi^\dag \hat{\psi})^{j-m'}}{\sqrt{(j-m')!}}
\end{equation}
which is homogeneous of degree $2j$ in $\xi_1^*$, $\xi_2^*$. Hence the expression

\begin{equation}
\frac{1}{\sqrt{(j+m')!}}~\frac{1}{\sqrt{(j-m')!}}~ (\frac{\partial}{\partial \xi_1^*})^{j+m}~(\frac{\partial}{\partial \xi_2^*})^{j-m}~ \frac{(\xi^\dag \psi)^{j+m'}}{\sqrt{(j+m')!}}~\frac{(\xi^\dag \hat{\psi})^{j-m'}}{\sqrt{(j-m')!}}  
\end{equation}
does not depend on $\xi$. Thus

\begin{eqnarray}
& & \frac{1}{\sqrt{(j+m')!}}~\frac{1}{\sqrt{(j-m')!}}~ (\frac{\partial}{\partial \xi_1^*})^{j+m}~(\frac{\partial}{\partial \xi_2^*})^{j-m}~ <\xi\mid U(\psi)\mid j,m'>= \nonumber  \\ & &~~~~= <\xi\mid  \frac{a_1^{j+m}}{\sqrt{(j+m)!}}~ \frac{a_2^{j-m}}{\sqrt{(j-m)!}}~U(\psi)~\mid j,m'>
\end{eqnarray}
does not depend on $\xi$. Hence we have

\begin{eqnarray}
& & <\xi\mid  ~ \frac{a_1^{j+m}}{\sqrt{(j+m)!}}~ \frac{a_2^{j-m}}{\sqrt{(j-m)!}}~U(\psi)~\mid j,m'>=  \nonumber  \\ & & ~~~~= <0\mid ~\frac{a_1^{j+m}}{\sqrt{(j+m)!}}~ \frac{a_2^{j-m}}{\sqrt{(j-m)!}}~U(\psi)~\mid j,m'>=D_{mm'}^j(\psi)
\end{eqnarray}
Thus

\begin{equation}
D_{mm'}^j(\psi)= \frac{1}{\sqrt{(j+m')!}}~\frac{1}{\sqrt{(j-m')!}}~ (\frac{\partial}{\partial \xi_1^*})^{j+m}~(\frac{\partial}{\partial \xi_2^*})^{j-m}~ <\xi\mid U(\psi)\mid j,m'>
\end{equation}
Using

\begin{equation}
x
<\xi\mid U(\psi)\mid j,m'> =  \frac{(\xi^\dag \psi)^{j+m'}}{\sqrt{(j+m')!}}~\frac{(\xi^\dag \hat{\psi})^{j-m'}}{\sqrt{(j-m')!}}  
\end{equation}
we finally get

\begin{eqnarray}
D_{mm'}^j(\psi)&=& \frac{1}{\sqrt{(j+m')!}}~\frac{1}{\sqrt{(j-m')!}}~ (\frac{\partial}{\partial \xi_1^*})^{j+m}~(\frac{\partial}{\partial \xi_2^*})^{j-m}~ \times \nonumber  \\
& &~~~~  \{\frac{(\xi^\dag \psi)^{j+m'}}{\sqrt{(j+m')!}}~\frac{(\xi^\dag \hat{\psi})^{j-m'}}{\sqrt{(j-m')!}}\}  
\end{eqnarray}

We note that $m$ numbers rows, $m'$ numbers columns. We have

\begin{equation}
D_{\frac{1}{2}~\frac{1}{2}}^{\frac{1}{2}}= \frac{\partial}{\partial \xi_1^*}~(\xi_1^* \psi_1 +\xi_2^* \psi_2)= \psi_1= e^{i(\alpha+\gamma)/2}~cos(\frac{\beta}{2})
\end{equation}

\begin{equation}
D_{\frac{1}{2}~-\frac{1}{2}}^{\frac{1}{2}}= \frac{\partial}{\partial \xi_1^*}~(-\xi_1^* \psi_2^* +\xi_2^* \psi_1^*)=-\psi_2^* ~~~~~~~~~~~~~~~~~~~~~
\end{equation}

\begin{equation}
D_{-\frac{1}{2}~\frac{1}{2}}^{\frac{1}{2}}= \frac{\partial}{\partial \xi_2^*}~(\xi_1^* \psi_1 +\xi_2^* \psi_2)= \psi_2= -e^{i(\gamma - \alpha)/2}~sin(\frac{\beta}{2})
\end{equation}

\begin{equation}
D_{-\frac{1}{2}~-\frac{1}{2}}^{\frac{1}{2}}= \frac{\partial}{\partial \xi_2^*}~(-\xi_1^* \psi_2^* +\xi_2^* \psi_1^*)= \psi_1^*~~~~~~~~~~~~~~~~~~~~~~
\end{equation}
so that

\begin{eqnarray}
D^{\frac{1}{2}}(\psi)&=& \pmatrix{D_{\frac{1}{2}~\frac{1}{2}}^{\frac{1}{2}} & D_{\frac{1}{2}~-\frac{1}{2}}^{\frac{1}{2}} \cr
D_{-\frac{1}{2}~\frac{1}{2}}^{\frac{1}{2}} & D_{-\frac{1}{2}~-\frac{1}{2}}^{\frac{1}{2}} \cr}= \pmatrix{ \psi_1 & -\psi_2^* \cr \psi_2 & \psi_1^* \cr}= \nonumber  \\ & =& e^{i\sigma_3 \frac{\alpha}{2}}~e^{i \sigma_2 \frac{\beta}{2}}~e^{i\sigma_3 \frac{\gamma}{2}}= \pmatrix{ e^{\frac{i}{2} (\alpha+\gamma)}~cos \frac{\beta}{2} & e^{-\frac{i}{2} (\gamma -\alpha)}~sin \frac{\beta}{2} \cr -e^{\frac{i}{2} (\gamma -\alpha)}~sin \frac{\beta}{2} & e^{-\frac{i}{2} (\alpha+\gamma)}~cos \frac{\beta}{2} \cr}  
\end{eqnarray}

Similarly

\begin{equation}
D^1(\psi)=\pmatrix{ \psi_1^2 & -\sqrt{2}  \psi_1 \psi_2^* & \psi_2^* \cr \sqrt{2} \psi_1 \psi_2 & \mid \psi_1\mid^2-\mid \psi_2\mid^2 & -\sqrt{2} \psi_1^* \psi_2^* \cr  \psi_2^2 & \sqrt{2} \psi_2 \psi_1^* & \psi_1^{*2} \cr}= e^{i\Sigma_3 \alpha}~e^{i\Sigma_2 \beta}~e^{i\Sigma_3 \gamma} 
\end{equation}
where $\Sigma_i$ are the $s=1$ (isotopic spin) matrices. They are

\begin{eqnarray}
\Sigma_1&=&\frac{1}{\sqrt{2}} \pmatrix{0 & 1 & 0 \cr 1 & 0 & 1 \cr 0 & 1 & 0 }~,~~~~\Sigma_2= \frac{1}{2i} \pmatrix{ 0 & \sqrt{2} & 0 \cr -\sqrt{2} & 0 & \sqrt{2} \cr 0 & -\sqrt{2} & 0 } \nonumber  \\
& & \Sigma_3= \pmatrix{ 1 & 0 & 0 \cr 0 & 0 & 0 \cr 0 & 0 & -1 \cr}
\end{eqnarray}
and we have

\begin{eqnarray}
\Sigma_+ = \Sigma_1 + i\Sigma_2 = \pmatrix{ 0 & \sqrt{2} & 0 \cr 0 & 0 & \sqrt{2} \cr 0 & 0 & 0 \cr}  \nonumber  \\
\Sigma_- = \Sigma_1 -i\Sigma_2 = \pmatrix{ 0 & 0 & 0 \cr \sqrt{2} & 0 & 0 \cr 0 & \sqrt{2} & 0 \cr}
\end{eqnarray}
These satisfy $[\Sigma_1, \Sigma_2]=i\Sigma_3$ and cyclic permutations. Now using angles $\phi=-\alpha$, $\theta=-\beta$ and $\psi=-\gamma$ we can write

\begin{eqnarray}
D^{\frac{1}{2}}(\psi)&=& \pmatrix{ \psi_1 & -\psi_2^* \cr \psi_2 & \psi_1^* \cr}=  e^{-i\sigma_3 \frac{\phi}{2}}~e^{-i \sigma_2 \frac{\theta}{2}}~e^{-i\sigma_3 \frac{\psi}{2}} \nonumber \\& & = \pmatrix{ e^{-\frac{i}{2} (\phi+\psi)}~cos \frac{\theta}{2} & -e^{-\frac{i}{2} (\phi -\psi)}~sin \frac{\beta}{2} \cr e^{\frac{i}{2} (\phi -\psi)}~sin \frac{\theta}{2} & e^{\frac{i}{2} (\phi+\psi)}~cos \frac{\theta}{2} \cr}  \label{eq:ellidort}
\end{eqnarray}
so that

\begin{equation}
\psi_1= cos \frac{\theta}{2}~e^{-i\frac{\phi+\psi}{2}} \label{eq:deellibes}
\end{equation}
and

\begin{equation}
\psi_2= sin \frac{\theta}{2}~e^{i \frac{\phi-\psi}{2}}  \label{eq:dellialti}
\end{equation}

Using these we can now write $D^1$ as

\begin{eqnarray}
D^1(\psi)= \pmatrix{ \frac{1+cos \theta}{2}~ e^{-i(\phi+\psi)} & -\frac{1}{\sqrt{2}}~ sin \theta ~e^{-i\phi} & \frac{1-cos \theta}{2}~ e^{-i(\phi-\psi)} \cr
\frac{1}{\sqrt{2}}~ sin \theta ~e^{-i\phi} & cos \theta & -\frac{1}{\sqrt{2}}~ sin \theta ~e^{i\phi} \cr
\frac{1-cos \theta}{2}~ e^{i(\phi+\psi)} & \frac{1}{\sqrt{2}}~ sin \theta ~e^{i\phi} & \frac{1+cos \theta}{2} ~e^{i(\phi+\psi)} \cr }
\end{eqnarray}

We have shown that we have formed $D$ functions with $\psi_1^{j+m} \psi_2^{j-m}$ in first columns. 

\section*{Euler Angles and $SU(2)/U(1)$ Coset Decomposition}

A group element for $SU(2)$ can be taken as

\begin{equation}
g_{\mbox{\boldmath$\omega$}} = A(\psi) = D^{\frac{1}{2}}(\psi)
\end{equation}
where
\begin{equation}
\psi = \left( \begin{array}{c}
\psi_1 \\  \psi_2  \end{array}  \right)
\end{equation}
with $\psi^\dag \psi =1$, or

\begin{equation}
\psi_1^* \psi_1 + \psi_2^* \psi_2 =1
\end{equation}

We can rewrite Eq.(\ref{eq:ellidort}) as

\begin{eqnarray}
D^{\frac{1}{2}}&=& \pmatrix{ \psi_1 & -\psi_2^* \cr \psi_2 & \psi_1^* \cr} = e^{-i \frac{\sigma_3}{2} ~\phi}~e^{-i \frac{\sigma_2}{2} ~\theta}~e^{-i \frac{\sigma_3}{2}~ \psi } \nonumber \\
& =& \pmatrix{ e^{-i \frac{\phi}{2}} & 0 \cr o & e^{i \frac{\phi}{2}} \cr} \pmatrix{ cos \frac{\theta}{2} & -sin \frac{\theta}{2} \cr sin \frac{\theta}{2} & cos \frac{\theta}{2} \cr} \pmatrix{ e^{-i \frac{\psi}{2}} & 0 \cr o & e^{i \frac{\psi}{2}} \cr} \nonumber \\
&=& \pmatrix{ cos \frac{\theta}{2}~e^{-i\frac{\phi +\psi}{2}} & -sin \frac{\theta}{2}~ e^{-i \frac{\phi-\psi}{2}} \cr
sin \frac{\theta}{2}~e^{i \frac{\phi-\psi}{2}} & cos \frac{\theta}{2}~e^{i \frac{\phi-\psi}{2}} \cr}  \label{eq:ellisekiz}
\end{eqnarray}

Let

\begin{equation}
z= \frac{\psi_2}{\psi_1} = tan \frac{\theta}{2}~e^{i\phi}
\end{equation}
and

\begin{equation}
\frac{\psi_1}{\mid \psi_1\mid } = e^{-i \frac{(\phi + \psi)}{2}} = e^{-i \frac{\eta}{2}}  \label{eq:altmisbir}
\end{equation}
where $z$ and $\eta$ are the parameters of the coset decomposition

\begin{equation}
SU(2)= (SU(2)/U(1))\times U(1)~.
\end{equation}
We can write $A=D^{\frac{1}{2}}$ as $A=BC$, where $C \in U(1)$ and $U(1)$ is the $U(1)$ subgroup of $SU(2)$ (locally isomorphic to $SO(2)$ subgroup of $SO(3)$). Now 

\begin{eqnarray}
D^{\frac{1}{2}}&=& \pmatrix{ \psi_1 & -\psi_2^* \cr \psi_2 & \psi_1^* \cr} \nonumber \\
&=& \pmatrix{ \psi_1 & -\psi_2^* \cr \psi_2 & \psi_1^* \cr} \pmatrix{ \frac{\mid \psi_1\mid }{\psi_1} &0 \cr 0 & \frac{\mid \psi_1\mid }{\psi_1^*} \cr} \pmatrix{ \frac{\mid \psi_1\mid }{\psi_1^*} &0 \cr 0 & \frac{\mid \psi_1\mid }{\psi_1} \cr} \nonumber  \\
&=& \pmatrix{ \mid \psi_1\mid  & -(\frac{\psi_2}{\psi_1})^*~\mid \psi_1\mid  \cr \frac{\psi_2}{\psi_1}~\mid \psi_1\mid  & \mid \psi_1\mid  \cr}
\pmatrix{ e^{-i \frac{\eta}{2}} & 0 \cr 0 & e^{i \frac{\eta}{2}} \cr} \nonumber \\
&=& \mid \psi_1 \mid \pmatrix{ 1 & -( \frac{\psi_2}{\psi_1})^* \cr \frac{\psi_2}{\psi_1} & 1 }
\pmatrix{ e^{-i \frac{\eta}{2}} & 0 \cr 0 & e^{i \frac{\eta}{2}} \cr} 
\end{eqnarray}

Since 
\begin{equation}
\mid \psi_1\mid^2 + \mid \psi_2\mid^2=\mid \psi_1\mid^2 (1+\mid \frac{\psi_2}{\psi_1}\mid^2)=1
\end{equation}
we have

\begin{equation}
\mid \psi_1\mid = \frac{1}{\sqrt{1+\mid z\mid^2}}
\end{equation}
Thus

\begin{equation}
D^{\frac{1}{2}} = \frac{1}{\sqrt{1+\mid z\mid^2}}~\pmatrix{ 1 & -z^* \cr z & 1 \cr} \pmatrix{ e^{-i \frac{\eta}{2}} & 0 \cr 0 & e^{i \frac{\eta}{2}} \cr}
\end{equation}

Now

\begin{equation}
B=\mid \psi_1 \mid \pmatrix{ 1 & -(\frac{\psi_2}{\psi_1})^* \cr (\frac{\psi_2}{\psi_1}) & 1 \cr} = \frac{1}{\sqrt{1+{\mid z \mid}^2}} \pmatrix{ 1 & -z^* \cr z & 1 \cr} 
\end{equation}
labels cosets $SU(2)/U(1)$. Therefore

\begin{equation}
A= \frac{1}{\sqrt{1+\mid z \mid^2}} \pmatrix{ 1 & -z^* \cr z & 1 \cr} \pmatrix{ \frac{\mid \psi_1 \mid}{\psi_1^*} &0 \cr 0 & \frac{\mid \psi_1 \mid}{\psi_1} \cr}
\end{equation}
with $\psi_1=z~\psi_2$ and $z$ is the coset label for $SU(2)/U(1)$. 

Under a rotation

\begin{equation}
\Omega= \pmatrix{ \omega_1 & \omega_2 \cr -\omega_2^* & \omega_1^* \cr}
\end{equation}
with

\begin{equation}
\mid \omega_1 \mid^2 + \mid \omega_2 \mid^2 =1
\end{equation}
we have

\begin{equation}
A' = \Omega A = \pmatrix{ \psi_1' & -\psi_2'^* \cr \psi_2' & \psi_1'^* \cr}
\end{equation}
or

\begin{equation}
\pmatrix{ \psi_1' \cr \psi_2' \cr} = \pmatrix{ \omega_1 & -\omega_2^* \cr \omega_2 & \omega_1^* \cr} \pmatrix{ \psi_1 \cr \psi_2 \cr}= \pmatrix{ \omega_1 \psi_1 -\omega_2^* \psi_2 \cr \omega_2 \psi_1 + \omega_1 \psi_2 \cr}
\end{equation}

We can write

\begin{equation}
A'= \frac{1}{\sqrt{1+\mid z' \mid^2}} \pmatrix{ 1 & -z'^* \cr z' & 1 \cr} \pmatrix{ \frac{\mid \psi_1' \mid}{\psi_1'^*} &0 \cr 0 & \frac{\mid \psi_1' \mid}{\psi_1'} \cr}
\end{equation}
with $\psi_2'=z' \psi_1'$.

Thus, for the coset label $z$ we have the transformation law

\begin{equation}
T_{\Omega} z = z' = \frac{\psi_2'}{\psi_1'} = \frac{\omega_2 \psi_1 + \omega_1^* \psi_2}{\omega_1 \psi_1 -\omega_2^* \psi_2} = \frac{\omega_2 + \omega_1^* z}{ \omega_1 - \omega_2^* z}
\end{equation}
which is a M\"obius transformation determined by $\Omega$.

We also have

\begin{equation}
\psi_1'= (\omega_1 -\omega_2^*z) \psi_1
\end{equation}

Thus far we have shown $A=A(z,\psi_1)$ and under $SU(2)$

\begin{equation}
A \rightarrow T_\Omega A= A(T_\Omega z, T_\Omega \psi_1)
\end{equation}

\begin{equation}
\Omega A= A(\frac{\omega_2 + \omega_1^* z}{\omega_1 - \omega_2^* z} , (\omega_1 -\omega_2^* z)\psi_1)
\end{equation}
where $z$ is the label for $SU(2)/U(1)$ and $\frac{\psi_1}{\mid \psi_1 \mid} = e^{-i\eta/2}$ where $\eta$ is the parameter for the $U(1)$ subgroup.

Using Eqs.(\ref{eq:ellisekiz}) and (\ref{eq:altmisbir}) we arrive at

\begin{equation}
A=BC = e^{-i \frac{\sigma_3}{2}\phi}~ e^{-i \frac{\sigma_2}{2}\theta}~ e^{-i \frac{\sigma_3}{2}\phi}~ e^{-i \frac{\sigma_3}{2}\eta}
\end{equation}
where

\begin{equation}
 B= e^{-i \frac{\sigma_3}{2}\phi}~ e^{-i \frac{\sigma_2}{2}\theta}~ e^{-i \frac{\sigma_3}{2}\phi}
\end{equation}
has the form

\begin{eqnarray}
B&=& \pmatrix{ e^{-i \frac{\phi}{2}} & 0 \cr 0 & e^{i \frac{\phi}{2}} \cr} \pmatrix{cos\frac{\theta}{2} & -sin \frac{\theta}{2} \cr sin\frac{\theta}{2} & cos\frac{\theta}{2} \cr} 
\pmatrix{ e^{i \frac{\phi}{2}} & 0 \cr 0 & e^{-i \frac{\phi}{2}} \cr} \nonumber  \\
&=& \pmatrix{cos\frac{\theta}{2} & -e^{-i\phi}sin \frac{\theta}{2} \cr e^{i\phi}sin\frac{\theta}{2} & cos\frac{\theta}{2} \cr} = cos\frac{\theta}{2} \pmatrix{ 1 & -e^{-i\phi} tan\frac{\theta}{2} \cr e^{i\phi} tan\frac{\theta}{2} & 1 \cr}
\end{eqnarray} 

If we put

\begin{equation}
z= -e^{-i\phi}~tan\frac{\theta}{2}   \label{eq:za}
\end{equation}
then

\begin{equation}
cos\frac{\theta}{2} =\frac{1}{\sqrt{1 + \mid z \mid^2}}
\end{equation}

On the other hand, if 

\begin{equation}
\frac{\psi_1}{\mid \psi_1 \mid} = e^{\frac{i}{2} (\psi+\phi)}  \label{eq:zb}
\end{equation}
we have

\begin{equation}
C= e^{-i \frac{\sigma_3}{2} (\psi+\phi)} = \pmatrix{ \frac{\mid \psi_1 \mid}{\psi_1^*} & 0 \cr 0 & \frac{\mid \psi_1 \mid }{\psi_1} \cr}
\end{equation}

Thus the formulas Eqs.(\ref{eq:za}) and (\ref{eq:zb}) give the relation between Euler angles and the coset decomposition with respect to $U(1)$. Note that $z$ involves only the two Euler angles $\phi$ and $\theta$.

For $j=1$ we find
\begin{eqnarray}
D^{(1)}(\psi)&=& \pmatrix{ \psi_1^2 & -\sqrt{2}  \psi_1 \psi_2^* & \psi_2^* \cr \sqrt{2} \psi_1 \psi_2 & \mid \psi_1\mid^2-\mid \psi_2\mid^2 & -\sqrt{2} \psi_1^* \psi_2^* \cr  \psi_2^2 & \sqrt{2} \psi_2 \psi_1^* & \psi_1^{*2} \cr} \times  \nonumber \\
& &~~\pmatrix{ \frac{\mid \psi_1\mid^2}{\psi_1^2} & 0 & 0 \cr 0 & 1 & 0 \cr 0 & 0 & \frac{\mid \psi_1\mid^2}{\psi_1^{*2}} \cr} 
\pmatrix{ \frac{\psi_1^2}{\mid \psi_1\mid^2} & 0 & 0 \cr 0 & 1 & 0 \cr 0 & 0 & \frac{\mid \psi_1\mid^{*2}}{\mid \psi_1\mid^2} \cr} \nonumber  \\
&=& \mid \psi_1\mid^2 \pmatrix{ 1 & -\sqrt{2} (\frac{\psi_2}{\psi_1})^* & (\frac{\psi_2}{\psi_1})^{*2} \cr
\sqrt{2} ~\frac{\psi_2}{\psi_1} & 1-\mid \frac{\psi_2}{\psi_1}\mid^2 & -\sqrt{2}~ \frac{\psi_2^*}{\psi_1^*} \cr
(\frac{\psi_2}{\psi_1})^2 & \sqrt{2}~ \frac{\psi_2}{\psi_1} & 1 \cr}  
\pmatrix{ e^{-i\eta} &0 &0 \cr 0 & 1 & 0 \cr 0 & 0 & e^{i\eta} \cr} 
\end{eqnarray}
or

\begin{eqnarray}
D^{(1)}= \frac{1}{1+\mid z\mid^2} \pmatrix{ 1 & -\sqrt{2}~ z^* & z^{*2} \cr \sqrt{2}~z & 1- z z^* & - \sqrt{2}~z^* \cr z^2 & \sqrt{2}~z & 1 \cr} \pmatrix{ e^{-i\eta} &0 &0 \cr 0 & 1 & 0 \cr 0 & 0 & e^{i\eta} \cr} 
\end{eqnarray}

In general the elements of 

\begin{equation}
(1+ \mid z\mid^2)^{2j}~D^{(j)}~e^{J_3 \eta} =Z
\end{equation}
are polynomials in $z$ and $z^*$. The first column is analytic and the last column is antianalytic.

\section*{Acknowledgments}
I thank Chia-Hsiung Tze and C. Devhand from whom I learned most all of this material. Discussions and working together with late professors Feza G\"ursey at Yale and Victor Ogievetsky at Rockefeller was of great joy and experience. 
I also thank Vladimir Akulov, Ramzi Khuri and Anatoly Pashnev for useful discussions.


\newpage

\begin{thebibliography}{99}
\bibitem{gurseyandtze} F. G\"ursey and C. Tze, {\em On the Role of Division, Jordan and Related Algebras in Particle Physics}. World Scientific (1996)
\bibitem{c1}See R. Penrose and W. Rindler, Spinors and Space-Time, Vol.1 and 2, (Cambridge University Press, 1984 and 1986); R.S. Ward and R.O. Wells, Jr., Twistor Geometry (Cambridge University Press);R.S. Ward, {\em Phys. Lett. \underline{61A} (1977) 81}.
\bibitem{c2} A. Galperin, E. Ivanov, V. Ogievetsky and E. Sokatchev, {\em Ann. Phys. \underline{185} (1988) 1}; and {\em Ann. Phys. \underline{185} (1988) 22.}
\bibitem{c2a}M. Evans, F. G\"ursey and V. Ogievetsky, {\em Phys. Rev. \underline{D47} (1993) 3496.}
\bibitem{c2b}C. Devchand and V. Ogievetsky, {\em Four-Dimensional Integrable Theories.} Proc. of the G\"ursey Memorial Conf. I, p. 169, Ed. G. Aktas and M. Serdaro\v{g}lu, Springer-Verlag (1995); hep-th/9410147.
\bibitem{c2c} F. G\"ursey and W. Jiang, {\em J. Math. Phys. \underline{32} (1991) 2365.}
\bibitem{c3} F. G\"ursey and C.H.-Tze, {\em Ann. Phys. \underline{128} (1980) 29.}
\bibitem{c4} Preservation of Kruskal form by Fueter transformations in euclidean gravity was first noted by F. G\"ursey and C. H.-Tze, {\em Lett. Math. Phys. \underline{8} (1984) 1187.}
\bibitem{c5} I.M. Gelfand and D.B. Fuchs, {\em Funct. Anal. Appl. \underline{2} (1968) 92 }; see also D.B. Fuchs, {\em Cohomology of Infinite Dimensional Lie Algebras} (Plenum, 1986).
\bibitem{c6} J.L. Cardy, {\em Nucl. Phys. \underline{B270} (1986) 186.}
\bibitem{c7} E.S. Fradkin and V.Y. Linetsky, {\em Phys. Lett. \underline{B 253} (1991) 97.}
\bibitem{c7a} E.S. Fradkin and M.Y. Palchik, {\em Conformal Field Theory in D-Dimensions} (World Scientific, 1993).
\bibitem{c8} V.P. Nair and J. Schiff, {\em Phys. Lett. \underline{246B} (1990) 423.}
\bibitem{c1a} R. Penrose, {\em Twistors, Particles, Strings and Links. The Interface of Mathematics and Particle Physics.} Eds. D.G. Quillen, G.B. Segal and S.T. Tsou, p.49 (Clarendon Press, Oxford 1990). 
\bibitem{c1b} M.A. Singer, {\em Twistors and Four-Dimensional Conformal Field Theory, ibid., p.181}
\bibitem{c1c} A.P. Hodges, {\em String Amplitudes and Twistors Diagrams: an Analogy, ibid., p.217}
\bibitem{c1d} M.A. Singer, {\em Comm. Math. Phys. \underline{133} (1990) 75.}
\bibitem{c1e} M.A. Atiyah, N.J. Hitchin and I.M. Singer, {\em Proc. Roy. Soc. London \underline{A362} (1978) 425.}
\bibitem{c1f} A.P. Hodges, R. Penrose and M.A. Singer, {\em Phys. Lett. \underline{B216} (1989) 48.}
\bibitem{martinec} E. Martinec, {\em "Geometrical Structure of M-theory."} arXiv:hep-th/9608017.
\bibitem{siegel} W. Siegel, {\em Phys. Rev. \underline{D52} (1995) 1042.} 
\bibitem{c9} H. Sugawara, {\em Phys. Rev. \underline{170} (1968) 1659.}
\bibitem{c10} C.M. Sommerfield, {\em Phys. Rev. \underline{176} (1968) 2019.} 
\bibitem{c11} M. Kaku, {\em Strings, Conformal Fields, and Topology} (Springer Verlag, 1991).
\bibitem{c12} C.-H. Gu, et. al. {\em Sci. Sin. \underline{21} (1978) 475.}
\bibitem{howe} See P.S. Howe, {\em "On Harmonic Superspace."} arXiv:hep-th/9812133.

\bibitem{maldacena} J. Maldacena, {\em Adv. Theor. Math. Phys. \underline{2} (1998) 231.}

\bibitem{ramzi} M.J. Duff, J.M. Evans, R.R, Khuri, J.X. Lu and R. Minasian, {\em Phys. Rev. \underline{B412} (1997) 281.}

\bibitem{ramond} P. Ramond, {\em "Algebraic Dreams."} arXiv:hep-th/0112261.
\bibitem{sultan} S. Catto, {\em Exceptional Projective Geometries and Internal Symmetries."} arXiv:hep-th/0302079.
\bibitem{kulkarni} R.S. Kulkarni, {\em J. Diff. Geom. \underline{13} (1978) 109.}
\bibitem{zucchini} R. Zucchini, {\em J. Geom. Phys. \underline{27} (1998) 113}; arXiv:gr-gc/9707048.
\bibitem{kruglov} S.I. Kruglov, {\em Int. J. Theor. Phys. \underline{41} (2002) 653}; arXiv:hep-th/0110251.

\end{thebibliography}
\end{document}